\documentclass[preprint]{emulateapj} 
\usepackage{hyperref}
\hypersetup{colorlinks,citecolor=Blue,linkcolor=Red,urlcolor=Blue}
\usepackage{amsmath}
\usepackage{empheq}
\usepackage[usenames,dvipsnames]{color}

\allowdisplaybreaks 
\hypersetup{colorlinks,citecolor=Blue,linkcolor=Red,urlcolor=Blue}
\usepackage[usenames,dvipsnames]{color}
\usepackage{amsmath,amsfonts,amssymb}
\usepackage{graphicx}
\usepackage{wasysym}
\usepackage{float}
\usepackage{url}
\usepackage{chngpage}
\usepackage{array}
\begin{document}

\title{TTV-determined Masses for Warm Jupiters and their Close Planetary Companions}  
\author{Dong-Hong Wu$^{1,2}$, Songhu Wang$^{3,4}$, Ji-Lin Zhou$^{1}$, Jason H. Steffen$^{2}$, Gregory Laughlin$^{3}$} 

\affil{$^{1}$School of Astronomy and Space Science and Key Laboratory of Modern Astronomy and Astrophysics in Ministry of Education, Nanjing University, Nanjing 210093, China} 
\affil{$^{2}$University of Nevada, Las Vegas, Department of Physics and Astronomy, 4505 S Maryland Pkwy, Box 454002, Las Vegas, NV 89154}
\affil{$^{3}$Department of Astronomy, Yale University, New Haven, CT 06511, USA}
\affil{$^{4}$\textit{51 Pegasi b} Fellow} 

\email{zhoujl@nju.edu.cn}
\email{song-hu.wang@yale.edu}

\begin{abstract} 

Although the formation and the properties of hot Jupiters (with orbital periods $P<10\,$d) have attracted a great deal of attention, the origins of warm Jupiters ($10<P<100\,$d) are less well-studied. Using a transit timing analysis, we present the orbital parameters of five planetary systems containing warm Jupiters, Kepler 30, Kepler 117, Kepler 302, Kepler 487 and Kepler 418. Three of them, Kepler-30 c($M_p=549.4{\pm{5.6}}M_{\oplus}$), Kepler-117 c($M_p=702{\pm{63}}M_{\oplus}$) and Kepler 302 c($M_p=933\pm 527M_{\oplus}$), are confirmed to be real warm Jupiters based on their mass. Insights drawn from the radius-temperature relationship lead to the inference that hot Jupiters and warm Jupiters can be roughly separated by ${T_{\rm eff,c}} =1123.7\pm3.3$ K. Also, ${T_{\rm eff,c}}$ provides a good separation for Jupiters with companion fraction consistent with zero($T_{\rm eff}>T_{\rm eff,c}$) and those with companion fraction significantly different from zero ($T_{\rm eff}<T_{\rm eff,c}$).

%Among the systems that we examined, only the planets in Kepler-30 (Kepler-30 b and Kepler-30 c) are near first order mean motion resonances. Moreover, dynamical simulations show that a $10\%$ increment in planetary mass or a $0.2\%$ period decrease of Kepler-30 c are sufficient to put the planet pair into resonance. 
%We also compare the stellar metallicity distribution of single Jupiter systems and systems in which a Jupiter-mass planet is associated with nearby companion planets, however, as there is limited data, no statistical conclusions can be made. %We also carry out a metallicity distribution comparison which shows that the metallicities of single-Jupiter systems are slightly larger than those of systems in which a Jupiter-mass planet is associated with nearby companion planets. This constitutes evidence that single Jupiters are more likely to arise as a consequence of a 'high-$e$' migration process. 

\end{abstract} 
%\maketitle
\section{Introduction}
%1. Hot Jupiter formation still most troublesome problem.
The identification of the main formation pathway for hot Jupiters(which we define as planets with orbital periods, $P$ $<$ 10 days, and masses, $M_{\rm p}$ $>$ $0.3 M_{\rm Jup}$) stands out as an unsolved problem. Although in-situ formation \citep{Bodenheimer2000} has been recently revived as a potential formation mechanism \citep{Batygin2016, Boley2016}, the conventional view holds that hot Jupiters could not have formed in their current locations, due to a lack of disk material and high temperatures near the host star \citep{Bell1997,Bodenheimer2000}.  It is generally believed that the observed population of short-period giant planets forms beyond the snow line (where the raw material is both abundant and cool) and experiences inward migration \citep{Lin1996}.

During the past 20 years, two competing migration narratives were established.  One pictures quiescent migration where giant planets exchange angular momentum with the surrounding disk and migrate towards the central star \citep{Goldreich1980,Lin1986,Lin1996,Masset2003}. The planets are envisioned to stop at the inner edge of the disk, where the disk gas and stellar spin co-rotate and where the disk is truncated with a magnetospheric cavity \citep{Shu1994,Ida2010}. The other mechanism involves migration following the generation of high orbital eccentricity. In this picture, the orbit of the giant planet is impulsively modified upon interaction with another object and is ultimately circularized by tidal interactions between the planet and the host star \citep{Nagasawa2008,Wu2011}. Frameworks that draw on this basic scenario include  planet-planet scattering \citep{Rasio1996,Nagasawa2008,Chatterjee2008,Beaug2012}, the Kozai-Lidov mechanism \citep{kozai1962,Wu2007,Naoz2012,Naoz2011,Chen2013,Anderson2016}, Secular chaos \citep{Wu2011} and excitation from passing stars \citep{Zakamska2004,Mart2015}. 

\begin{figure}
\vspace{0cm}\hspace{0cm}
\centering
\includegraphics[width=\columnwidth]{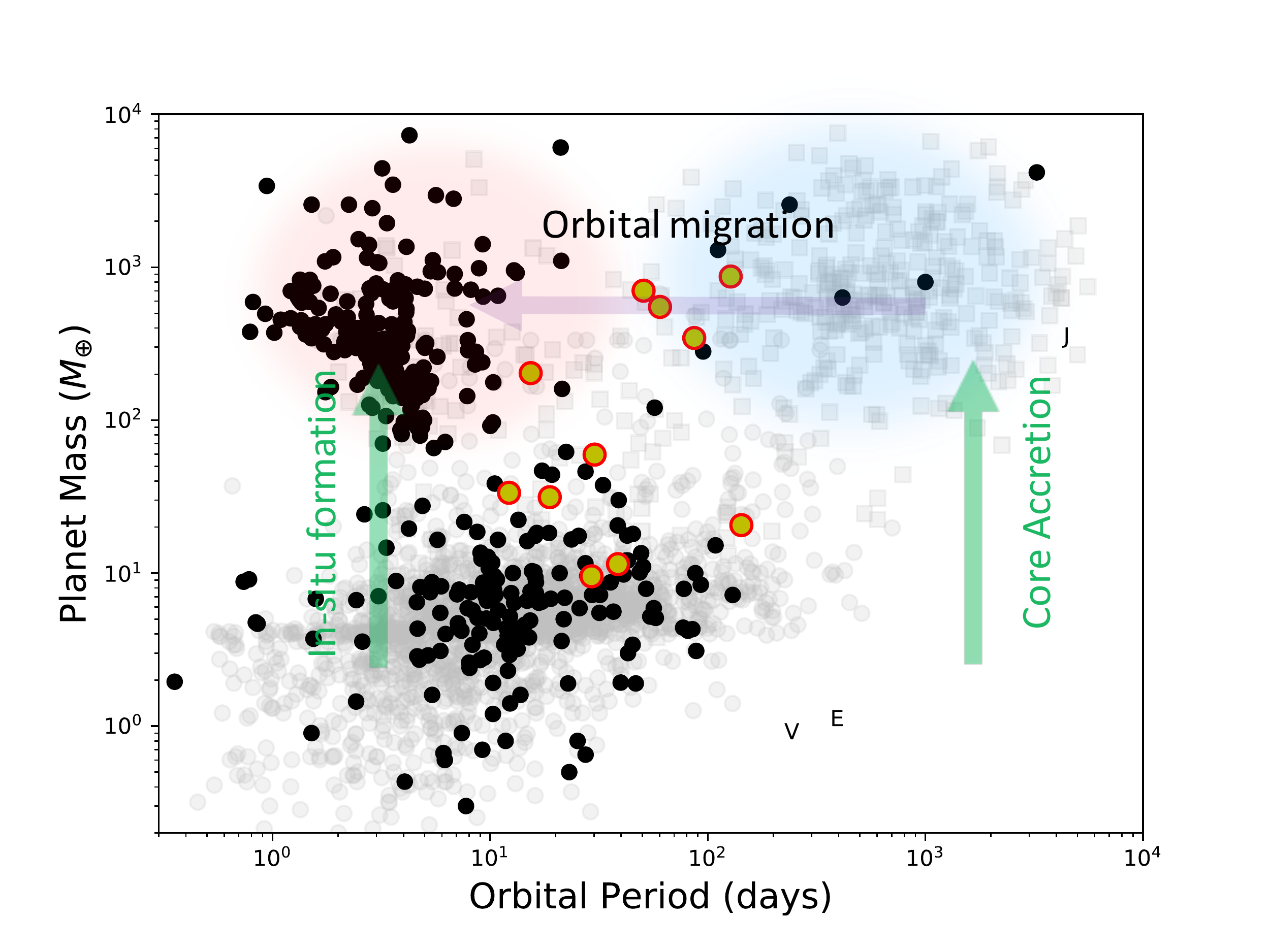}
\caption{Mass-period diagram for the population of known exoplanets. Transiting planets with precise mass measurements -- obtained using either the transit timing or the radial velocity technique -- are shown as solid black circles. Non-transiting planets with measured $M\sin(i)$ are shown as transparent squares, and planets with masses obtained from the mass-radius relationship are shown in transparent dots. Planets with masses calculated in this paper are shown as yellow dots with red circles. Planets with transit timing-determined masses are from \citet{Hadden2017}, and other planets are from exoplanets.org.}
\label{fig1}
\end{figure}

The relative contributions from each mechanism to the overall distribution of hot Jupiters remains unclear. There has been a long-standing suggestion that the lack of nearby planetary companions in hot Jupiter systems \citep{Steffen2012,Huang2016,Yong2017} can be taken as evidence that hot Jupiters owe their current orbits to violent dynamical histories---although the observational biases are not fully understood \citep{Wang2017}.  The WASP-47 planetary system \citep{Hellier2012}, remains the only known counter-example -- this system contains a hot Jupiter that is closely accompanied by two nearby, much smaller planetary companions \citep{Becker2015,Neveu2016}.  As such, it merits particular scrutiny and has already been the subject of considerable additional study \citep{Sinukoff2017,Weiss2017}. 

As with other short-period planets, warm Jupiters are frequently held as unlikely to have formed in the locations where they end up. Their migration histories, however, are less straightforward. They orbit too far from their host stars for tidal circularization to have been effective \citep{Dong2014}, and halting their migration using the magnetospheric inner disk truncation mechanism does not apply \citep{Ida2010}.

Although observational biases may have a contribution, warm Jupiters do seem to occupy a period ``valley'' between hot Jupiters and the more distant (and more numerous) population of massive planets (Figure~\ref{fig1}), which raises questions about their progenitors.  Are they a separate population of giant planets?  Or, they share a similar formation process to hot or cold Jupiters?  Do they migrate to their current location?  Or, do they form locally through the accretion \citep{Chiang2013,Lee2014,Batygin2016}.  

So far, 11 out of the 27 warm Jupiter systems in the Kepler data set have been determined to harbor nearby companions \citep{Huang2016}.  Unlike WASP-47, however, the properties of these systems have not yet been well studied. For example, the warm Jupiter systems listed in \citet{Huang2016} were identified by their radii rather than their masses (planets with $R_p$ $>$ $8R_{\oplus}$ were classified as warm Jupiters). Moreover, planets with radii similar to Jupiter sometimes turn out to be super-Earths, such as Kepler-9 b and Kepler-9 c \citep{Wangwu2017}. In this paper, we aim to determine the masses of the `warm Jupiter' systems, thereby helping to further delineate this interesting population.

\begin{figure}
\vspace{0cm}\hspace{0cm}
\centering
\includegraphics[width=\columnwidth]{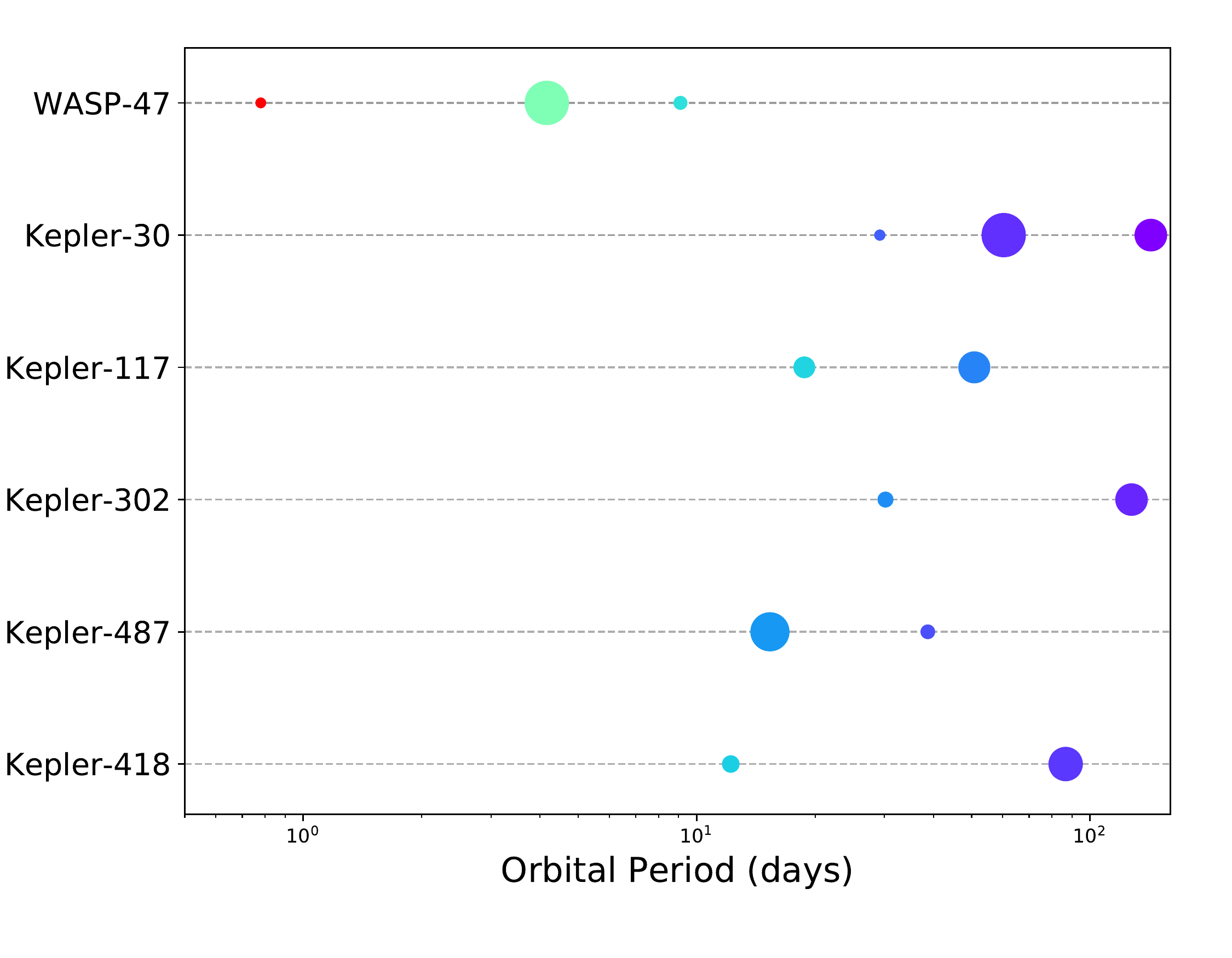}
\caption{Orbital architectures of six planetary systems considered in this paper. The sizes of the dots are proportional to the radii of the planets. Planets are colored according to equilibrium temperature ranging from 336 K (purple) to 2219 K (red).}
\label{fig2}
\end{figure}

Transit Timing Variations (TTVs) have been used to measure the masses and the eccentricities of planets orbiting stars that are too faint to support high-precision Doppler velocity measurements \citep{Agol2005,Holman2005,Lithwick-12012,JH2016,Hadden2017,Wangwu2017}. In this paper, we calculate the system parameters of five warm Jupiter systems with nearby companions -- Kepler-30, Kepler-117, Kepler-302, Kepler-487 and Kepler-418 -- that are listed in Table 4 of \citet{Huang2016}(Figure \ref{fig2}). These systems each have a full 17-Quarter time series of Kepler data and all of them show measurable TTV signals.

The structure of this paper is as follows. In Section 2, we describe the method we use to obtain planetary parameters from the TTV data. In section 3, we discuss the properties of the warm Jupiter systems we have analyzed. And we summarize our results in Section 4.

\section{Methodology}

\subsection{Target selection}

Among the 27 warm Jupiter systems listed in \citet{Huang2016}, 11  have close-in companions. We selected 5 of these systems for our study, including Kepler-30, Kepler-117, Kepler-302, Kepler-487 and Kepler-418. Among the remaining systems, Kepler-89 is excluded because its planet Kepler-89 d was shown by \citet{Hadden2017} to be a super Earth instead of a warm Jupiter. Kepler-148 is excluded because the period ratio  between its warm Jupiter candidate KOI-398.01 and the inner two planets is larger than 12, and the TTVs of the outer warm Jupiter are not readily explained by the dynamical interaction with the inner two planets. Kepler-46, Kepler-289 and Kepler-419 are excluded because they have been studied in detail in previous works \citep{Nesvorny2012,Schimitt2014,Dawson2012,Dawson2014}.  Finally, KOI-6132 was excluded due to its incomplete record of transit times in the \citet{Holczer2016} catalog.

\subsection{Method}
 We carry out a differential evolution Markov Chain Monte Carlo-based (DEMCMC, \citealt{Ter2006}) analysis to infer the orbital parameters of the planets based on their transit mid-times and uncertainties from the \textit{full} 17 quarters of Kepler data \citep{Holczer2016}.  The free parameters considered by the DEMCMC fit are {$P$, $e$, $i$, $\omega$+$M_{0}$, $\omega$-$M_{0}$, $\Delta\Omega$, and $M_{\rm P}$}, where $P$ is the orbital period, $e$ is the eccentricity,  $i$ is the orbital inclination, $\omega$ is the argument of periastron, $M_{0}$ is the initial mean anomaly, $\Delta\Omega$ is the difference of the ascending nodes between two planets, (We fix the $\Omega$ of one planet to be 0, the difference of $\Omega$ between other planets and the fixed planet is then $\Delta\Omega$) and $M_{\rm P}$ is the planetary mass.  The central stellar mass $M_\star$ and radius $R_\star$ of each planetary system are fixed to the median values given in the Q1-Q16 KOI catalogue \citep{Mullally2015}.  We use the TTVFast code developed by \citet{Deck2014} to compute the transit mid-times.
 
Our priors on $P$ are obtained from \citet{Holczer2016} with uncertainties of $\pm0.01$ d. Priors on $e$ are normally distributed with median value of 0.04 and standard deviation of 0.02, in keeping with \citet{Xie2016}'s demonstration that eccentricity for multiple planet systems is concentrated around $e=0.04$. To avoid negative eccentricities, the minimal prior of the eccentricity is set to be 0.001. Priors on $\omega$+$M_{0}$ are estimated according to the period and the transit time, and priors on $\omega$-$M_{0}$ are randomly distributed between $-360^{\circ}$ and $360^{\circ}$. Priors on $\Delta\Omega$ are randomly distributed between $1^{\circ}$ and $3^{\circ}$(that is $\Omega$ of one planet is fixed to be 0, while priors on $\Omega$ of other planets are randomly distributed between $1^{\circ}$ and $3^{\circ}$, as previous study \citet{fang2012} indicates that the mutual inclinations between planets in multiple planet systems are very small.) and priors on $i$ are randomly distributed between $90^{\circ}-i_{\rm c}<=i<=90^{\circ}+i_{\rm c}$, $i_{\rm c}=\arctan((R_\star+R_p)/a)$. $i_{\rm c}$ is chosen to enable the transit of the planet. As for the masses, we assume a normally distributed prior with median value of 1 $M_{\rm Jup}$ and the standard deviation of 0.5 $M_{\rm Jup}$ for the Jupiter-sized planets. For smaller planets, we estimate the planetary mass according to the relation $M_p=(R_p/R_{\oplus})^{2.06}$ from \citet{lissauer2011}, and the priors on mass are normally distributed with median value as the estimated mass and standard deviation as half of the estimated mass.
 
 \begin{figure}
\vspace{0cm}\hspace{0cm}
\centering
\includegraphics[width=\columnwidth]{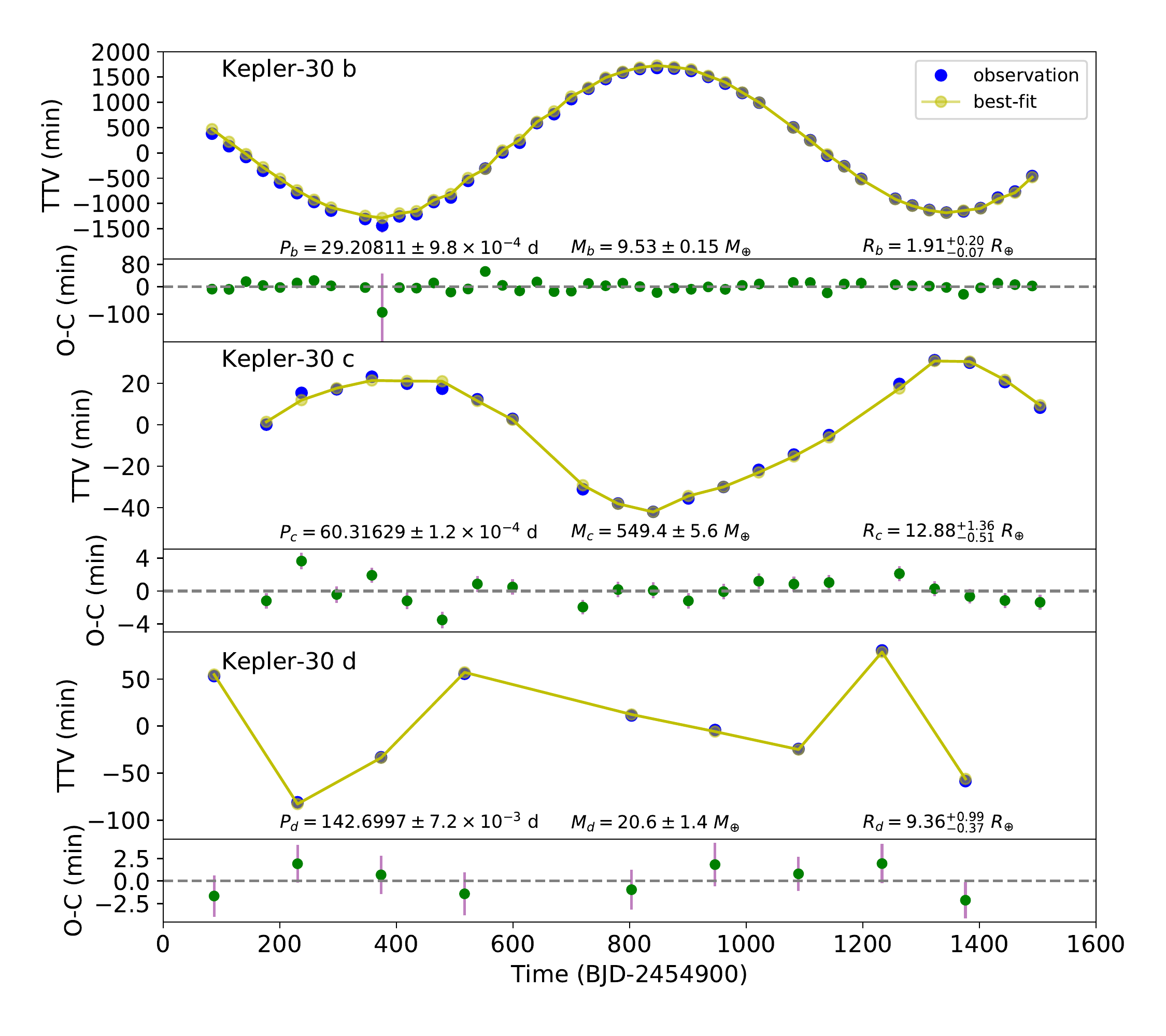}
\caption{Transit timing variations (blue dots) for the planets in the Kepler-30 system compared to our best fit (open yellow-green circles) obtained using the full Kepler data set.  The residuals of TTVs for each planet are shown as green dots below the corresponding panels.  As the errors are smaller than the symbols, they are invisible in the transit timing variations plots.}
\label{fig3}
\end{figure}

 We ran 40 parallel DEMCMC chains for two-planet systems and 60 parallel DEMCMC chains for three-planet systems, the number of chains are chosen according to \citet{Nelson2014}, who recommended $N_{\rm chains}$ $\sim$ $3N_{\rm dim}$, where $N_{\rm chains}$ is the number of DEMCMC chains and $N_{\rm dim}$ is the number of fitting parameters. The MCMC chains are run until the $\chi^2_{\rm red}$ of at least $80\%$ chains are stationary and the $\hat{R}$ statistics of all parameters in these chains were below 1.1 \citep{Brooks1998}. We record the parameters for each chain every 1000 generations for at least $4\times10^7$ generations. We discard the first 25\% ($1\times10^7$) of the samples in each chain to account for the burn-in phase, thereby reducing the risk of spurious parameter correlations. The step-size between two states in a chain is automatically adjusted to guarantee an acceptance rate between 0.2-0.3 \citep{Ford2006}. 
 
\begin{table*}[htbp]
\begin{center}
\tiny
%\resizebox{\textwidth}{
\caption{\bf The best-fit planetary parameters and the assumed stellar parameters, the reference epoch is $T_{0}=2454900.0 \,{\rm BJD_{TDB}}$}
\label{tab1}
\begin{adjustwidth}{-0.7cm}{0cm}
\begin{tabular}{ccccccccccc}
\tableline\tableline
Name&$M_*$($M_{\odot}$)&$R_*$($R_{\odot}$)&$R_p$($R_{\oplus}$)&$M_{\rm P}$($M_{\oplus}$)&P (day)&$e$&$i$($^\circ$)&$\omega$($^\circ$)&$\Omega$($^\circ$)&$M_{0}$($^\circ$)\\\hline
Kepler-30 b&1&0.95&$1.91^{+0.20}_{-0.07}$&$9.53{\pm0.15}$&$29.20811{\pm9.8\times10^{-4}}$&$0.07641{\pm2.7\times10^{-4}}$&$90.00{\pm0.77}$&$40.36{\pm0.27}$&$1.09{\pm0.80}$&$330.85{\pm0.24}$\\
Kepler-30 c&..&..&$12.88^{+1.36}_{-0.51}$&$549.4{\pm5.6}$&$60.31629{\pm1.2\times10^{-4}}$&$0.01148{\pm7.3\times10^{-4}}$&$89.99{\pm0.49}$&$315.2{\pm3.1}$&0&$262.8{\pm3.2}$\\
Kepler-30 d&..&..&$9.36^{+0.99}_{-0.37}$&$20.6{\pm1.4}$&$142.6997{\pm7.2\times10^{-3}}$&$0.0304{\pm2.0\times10^{-3}}$&$90.02{\pm0.27}$&
$198.5{\pm3.0}$&$2.5{\pm1.5}$&$231.1{\pm3.1}$\\
Kepler-117 b&1.205&1.183&$6.04^{+3.66}_{-0.75}$&$31.4{\pm8.6}$&$18.7767{\pm1.8\times10^{-3}}$&$0.0535{\pm6.2\times10^{-3}}$&$90.0{\pm1.2}$&$255.5{\pm4.4}$&$9.6{\pm2.3}$&$288.6{\pm2.6}$\\
Kepler-117 c&..&..&$9.16^{+5.55}_{-1.14}$&$702{\pm63}$&$50.78328{\pm4.5\times10^{-4}}$&$0.0344{\pm5.1\times10^{-3}}$&$89.97{\pm0.59}$&$300.5{\pm7.8}$&0&$120.8{\pm8.3}$\\
Kepler-302 b&0.966&0.833&$2.88^{+1.15}_{-0.24}$&$60{\pm49}$&$30.197{\pm0.012}$&$0.198{\pm0.065}$&$89.99{\pm0.73}$&$278{\pm44}$&$-6{\pm26}$&$97{\pm51}$\\
Kepler-302 c&..&..&$9.32^{+3.76}_{-0.77}$&$933{\pm527}$&$127.45{\pm0.15}$&$0.038{\pm0.037}$&$90.05{\pm0.26}$&$263{\pm95}$&$0$&$118{\pm92}$\\
Kepler-487 b&0.903&0.87&$11.31^{+4.78}_{-1.04}$&$203{\pm197}$&$15.35587{\pm2.1\times10^{-4}}$&$0.038{\pm0.034}$&$89.9{\pm1.3}$&$302{\pm106}$&$0$&$67{\pm88}$\\
Kepler-487 c&..&..&$2.62^{+1.11}_{-0.24}$&$11.5{\pm6.5}$&$38.676{\pm0.029}$&$0.03{\pm0.02}$&$89.97{\pm0.63}$&$254{\pm99}$&$0.2{\pm4.2}$&$195{\pm79}$\\

Kepler-418 b&1.039&1.043&$9.88^{+4.67}_{-1.02}$&$241{\pm213}$&$86.75{\pm0.07}$&$0.03{\pm0.03}$&$90.01{\pm0.43}$&$146{\pm106}$&$0$&$203{\pm99}$\\
KOI 1089.02&..&..&$4.75^{+2.25}_{-0.49}$&$138{\pm69}$&$12.21559{\pm6.8\times10^{-4}}$&$0.087{\pm0.098}$&$90.3{\pm1.4}$&$168{\pm98}$&$25{\pm42}$&$214{\pm100}$\\\hline

\end{tabular}
\end{adjustwidth}
\end{center}
\end{table*}

\begin{figure}
\vspace{0cm}\hspace{0cm}
\centering
\includegraphics[width=\columnwidth]{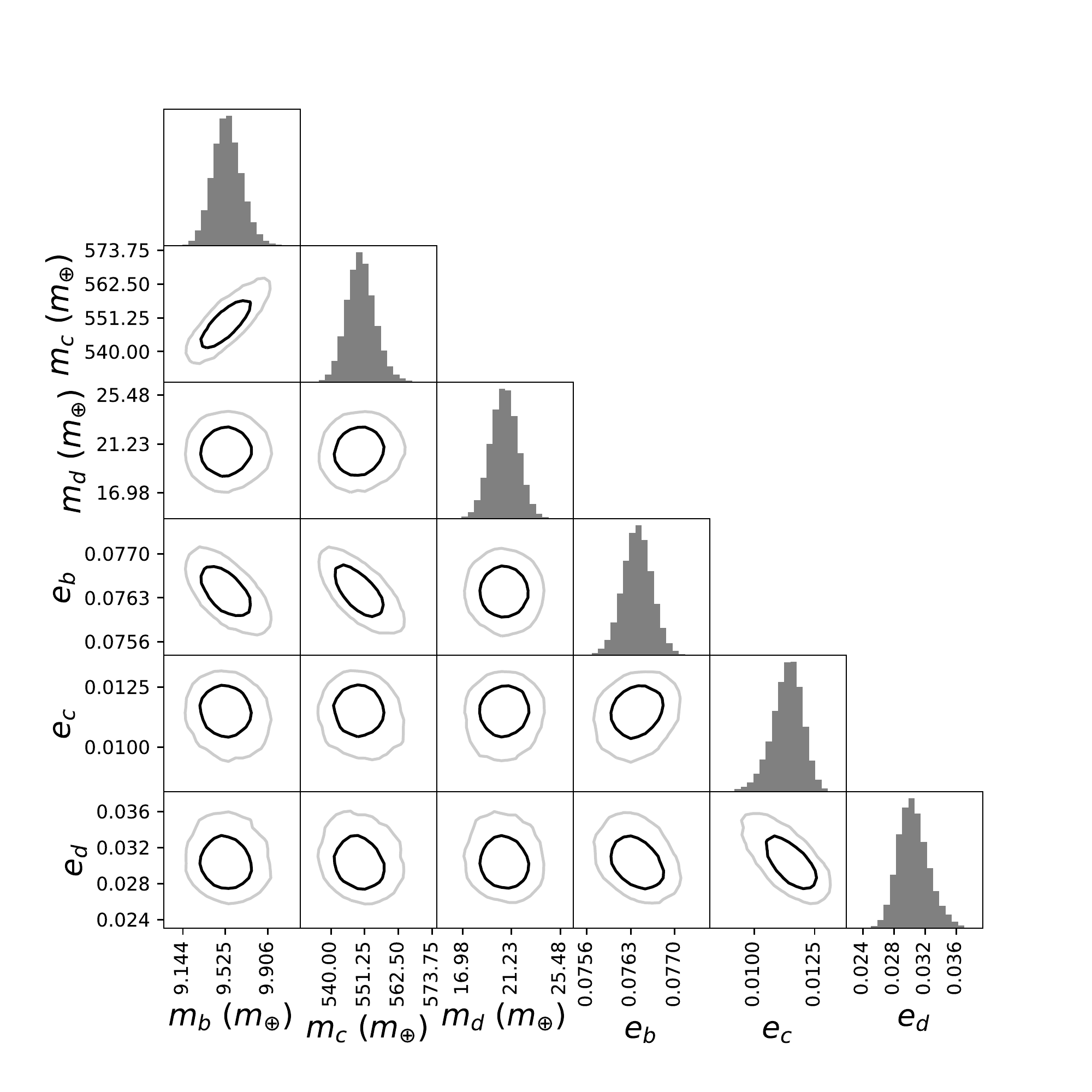}
\caption{Combined posterior mass and eccentricity distributions for the planets in Kepler-30. The dark and gray circles indicate the $68.3\%$ and $95.5\%$ confidence regions, respectively. Histograms show the marginalized mass and eccentricity posterior distributions from the DEMCMC.}
\label{fig4}
\end{figure}

\section{Results and Discussion}

\subsection{DEMCMC results}

\begin{figure}
\vspace{0cm}\hspace{0cm}
\centering
\includegraphics[width=\columnwidth]{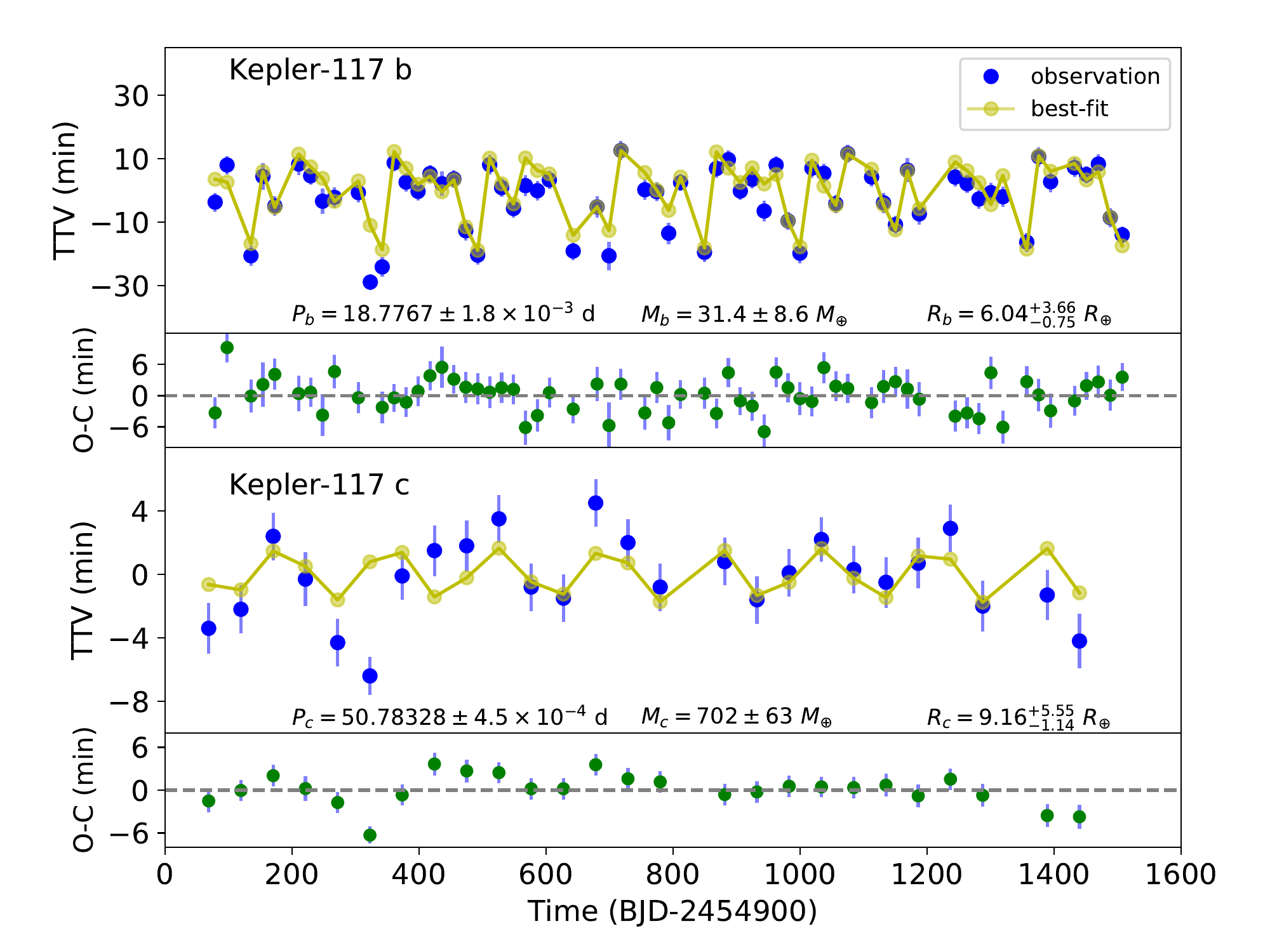}
\caption{Transit timing variations (blue dots) of the planets in Kepler-117 compared to our best fit (connected open yellow-green circles) obtained using the full Kepler data set. The residuals of TTVs for each planet are shown as green dots below the corresponding panels. }
\label{fig5}
\end{figure}

\begin{figure}
\vspace{0cm}\hspace{0cm}
\centering
\includegraphics[width=\columnwidth]{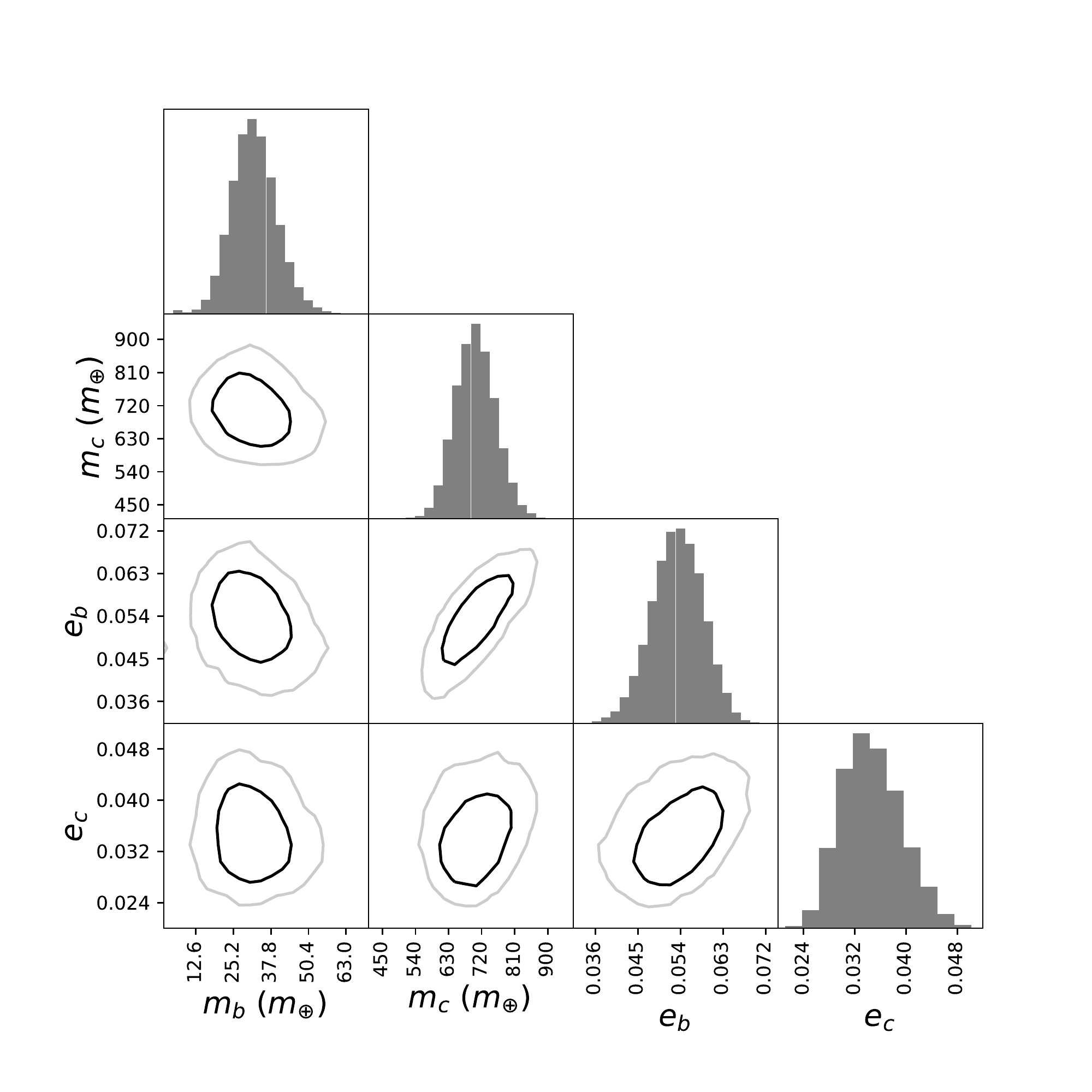}
\caption{Combined posterior mass and eccentricity distributions of the planets in Kepler-117. See Figure \ref{fig4} for description.}
\label{fig6}
\end{figure}

\begin{figure}
\vspace{0cm}\hspace{0cm}
\centering
\includegraphics[width=\columnwidth]{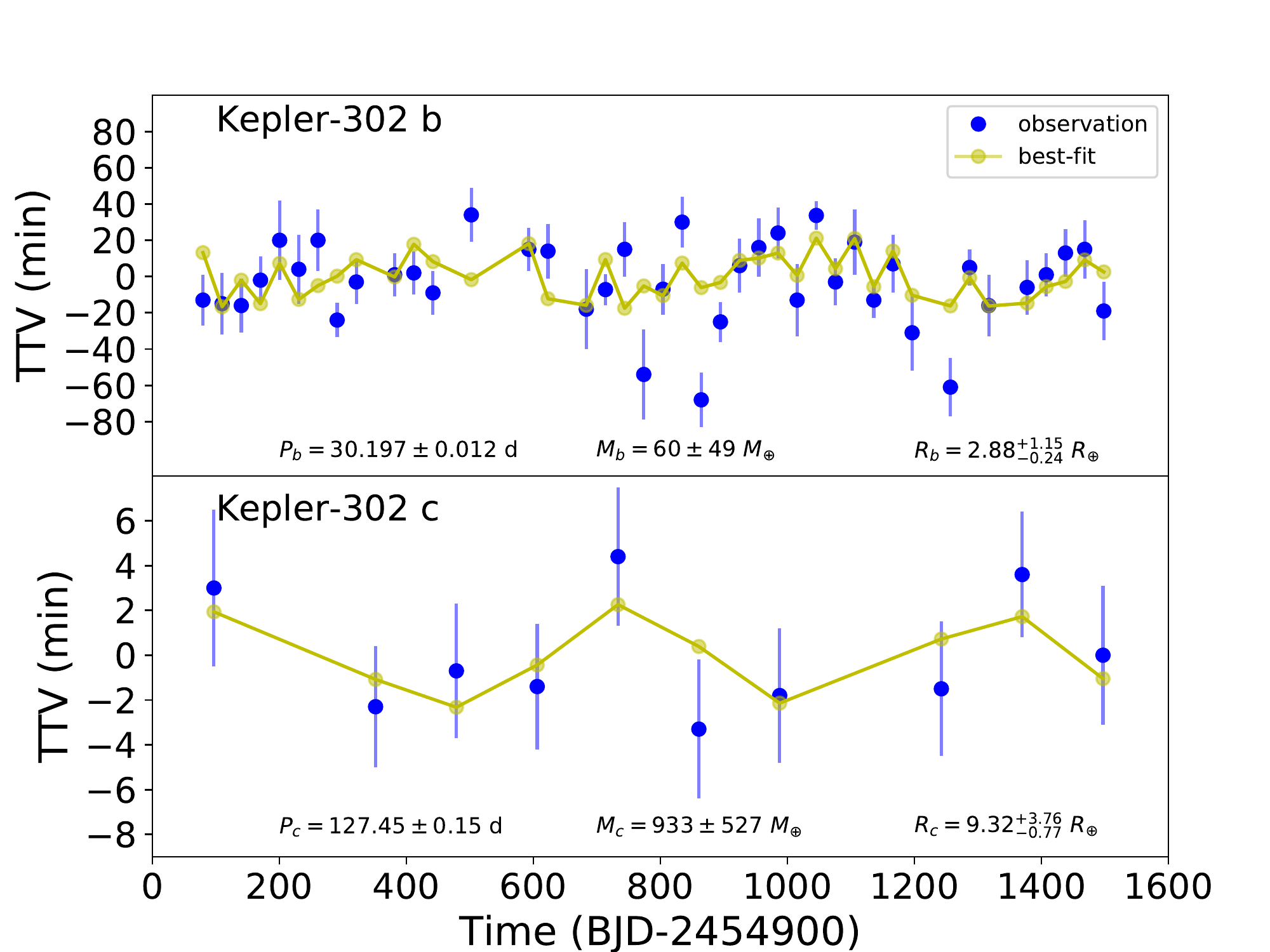}
\caption{Transit timing variations (blue dots) of planets in the Kepler-302 system compared to our best fit (connected open yellow-green circles) obtained using the full Kepler data set.}
\label{fig7}
\end{figure}

\begin{figure}
\vspace{0cm}\hspace{0cm}
\centering
\includegraphics[width=\columnwidth]{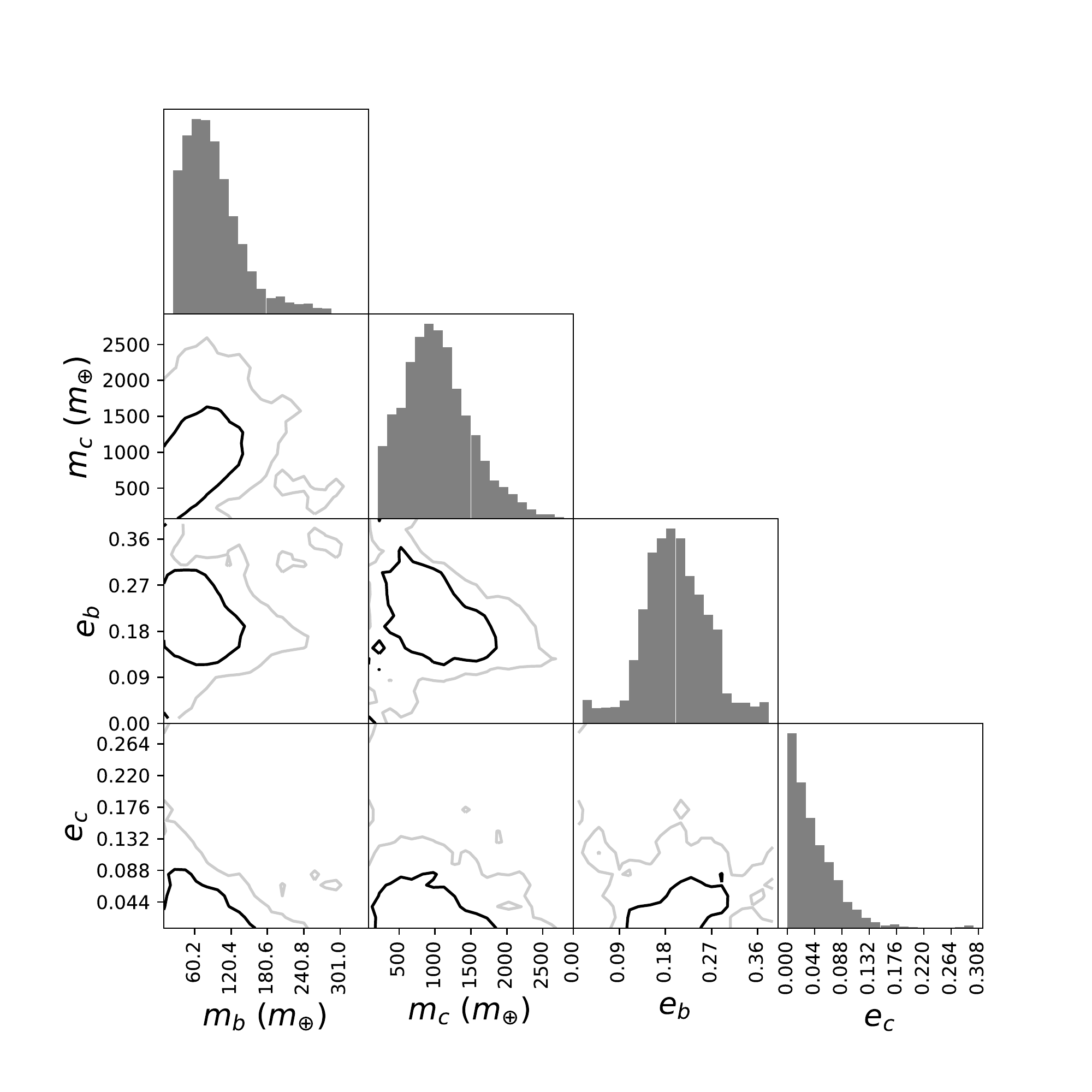}
\caption{Combined posterior mass and eccentricity distributions for the planets in the Kepler-302 system. See Figure \ref{fig4} for description.}
\label{fig8}
\end{figure}

\begin{figure}
\vspace{0cm}\hspace{0cm}
\centering
\includegraphics[width=\columnwidth]{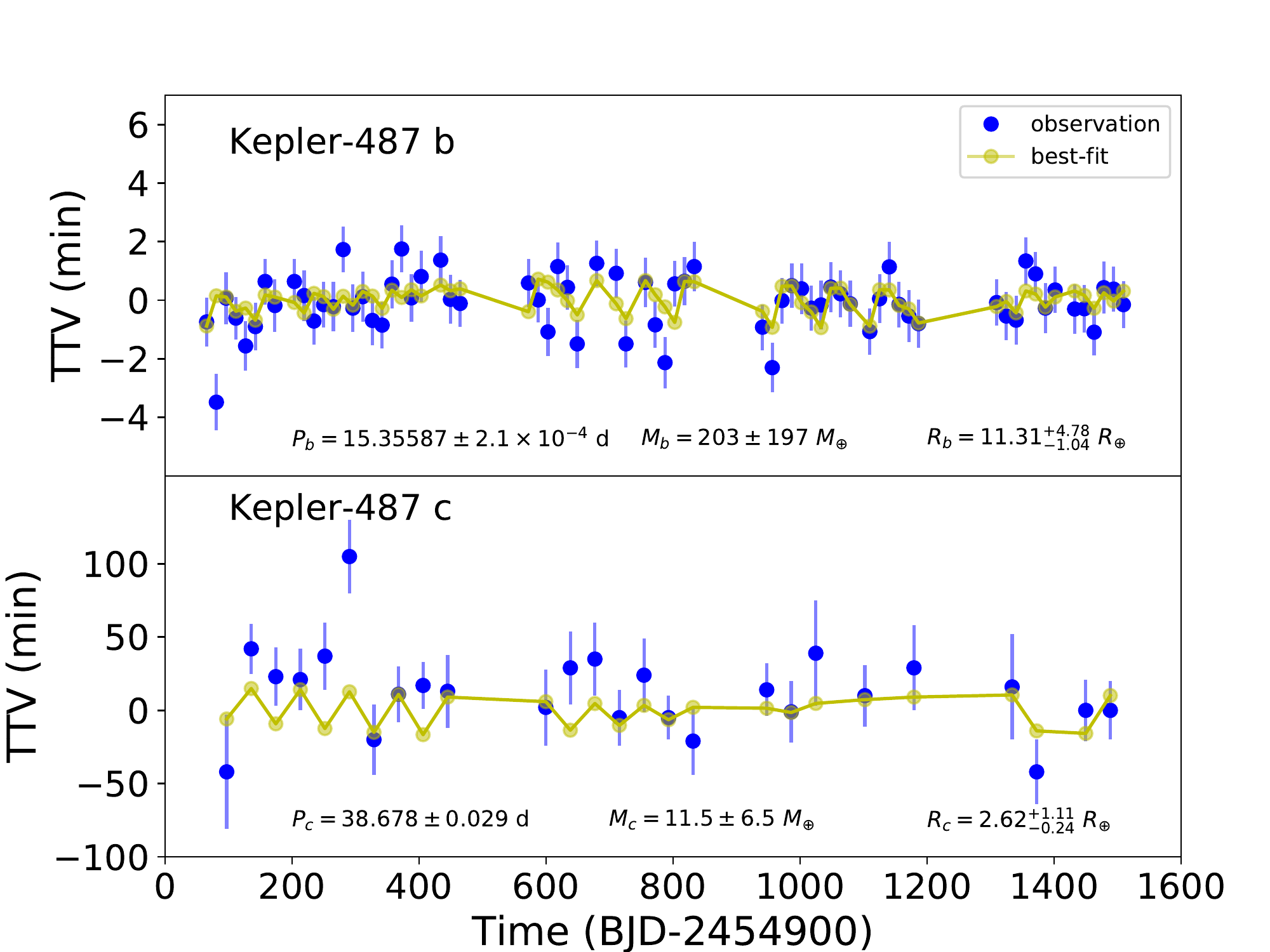}
\caption{Transit timing variations (blue dots) of the planets in the Kepler-487 system compared to our best fit (connected open yellow-green circles) obtained using the full Kepler data set.}
\label{fig9}
\end{figure}

\begin{figure}
\vspace{0cm}\hspace{0cm}
\centering
\includegraphics[width=\columnwidth]{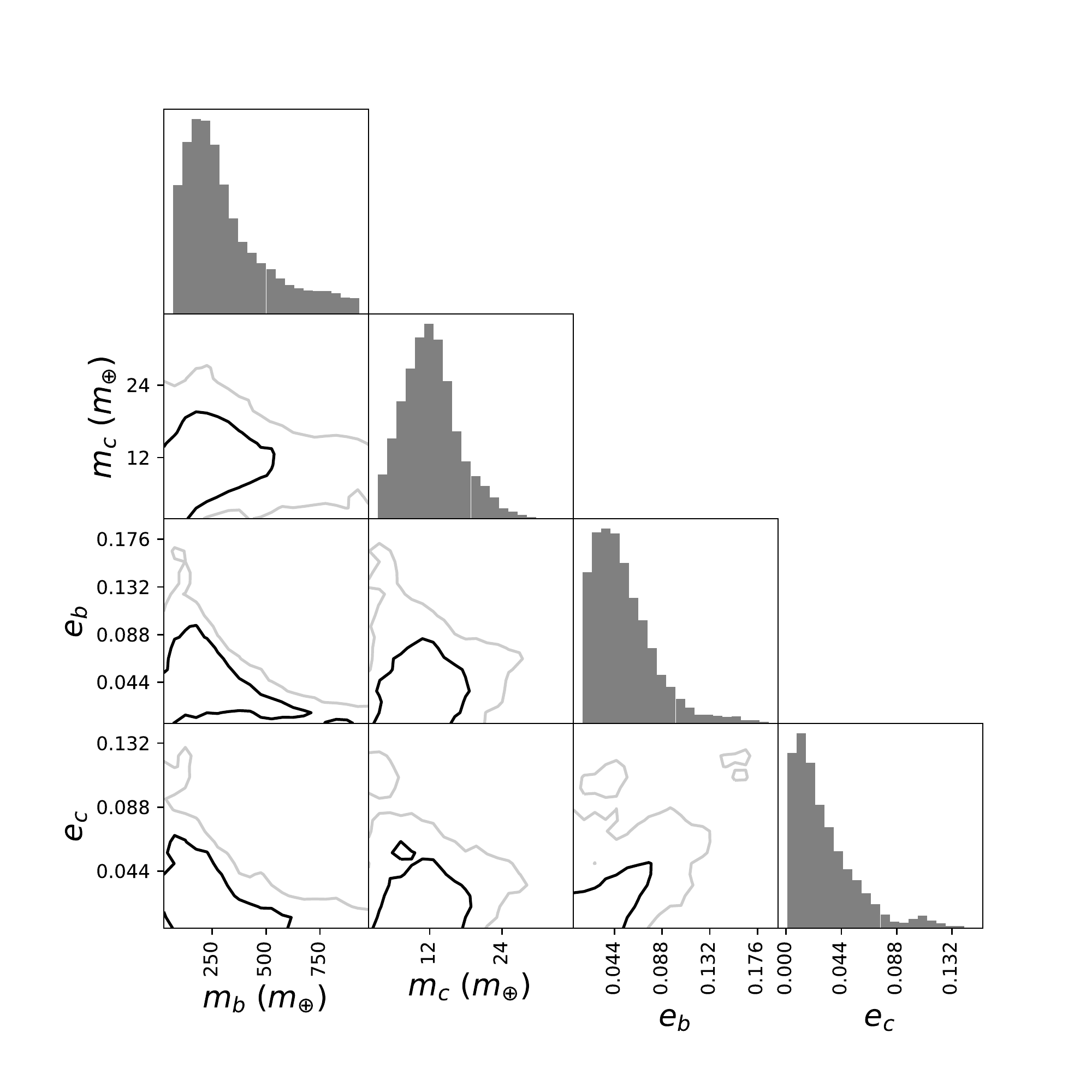}
\caption{Combined posterior mass and eccentricity distributions of the planets in the Kepler-487 system. See Figure \ref{fig4} for description.}
\label{fig10}
\end{figure}

\begin{figure}
\vspace{0cm}\hspace{0cm}
\centering
\includegraphics[width=\columnwidth]{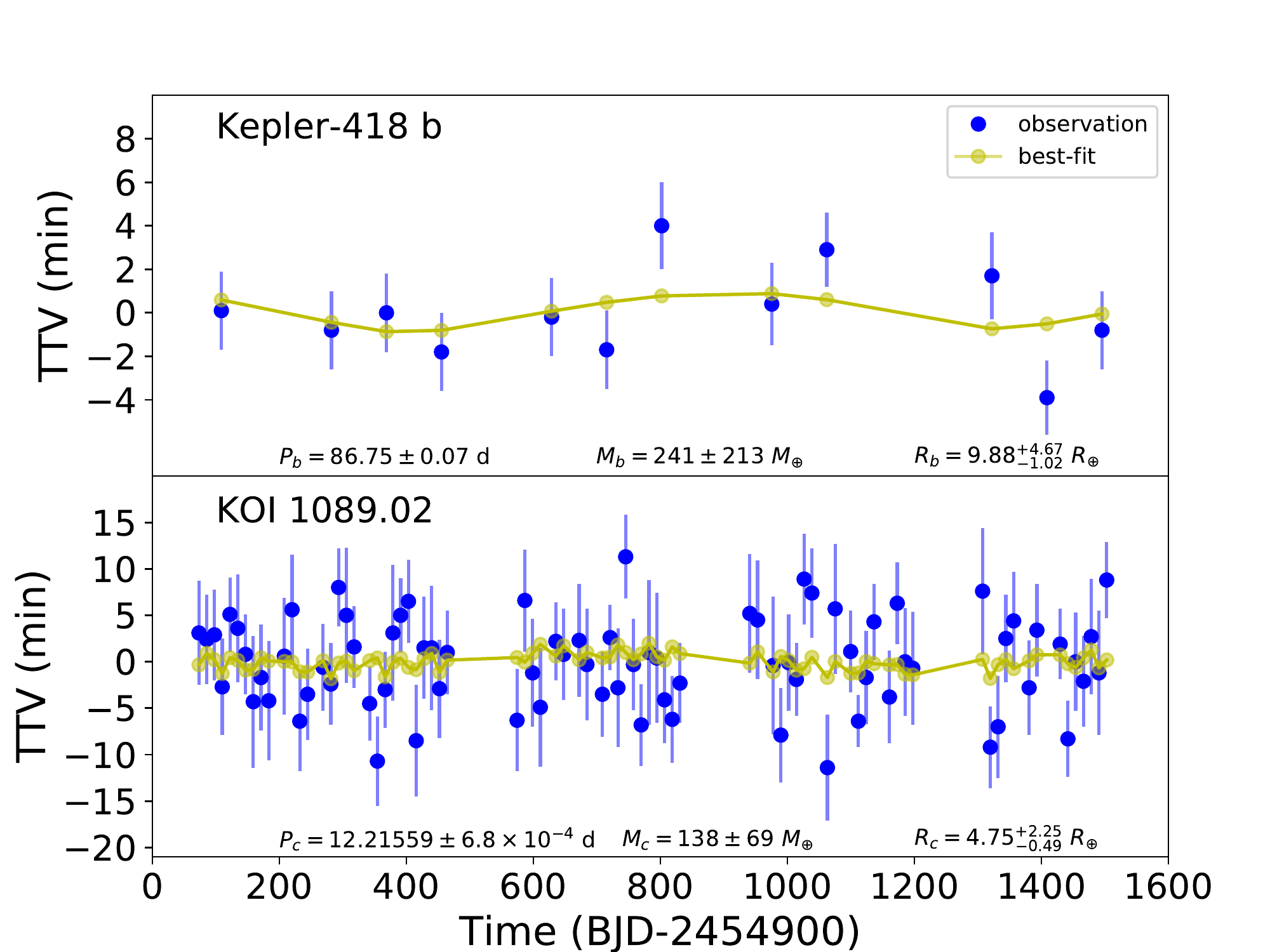}
\caption{Transit timing variations (blue dots) for the planets in the Kepler-418 system compared to our best fit (connected open yellow-green circles) obtained using the full Kepler data set.}
\label{fig11}
\end{figure}

\begin{figure}
\vspace{0cm}\hspace{0cm}
\centering
\includegraphics[width=\columnwidth]{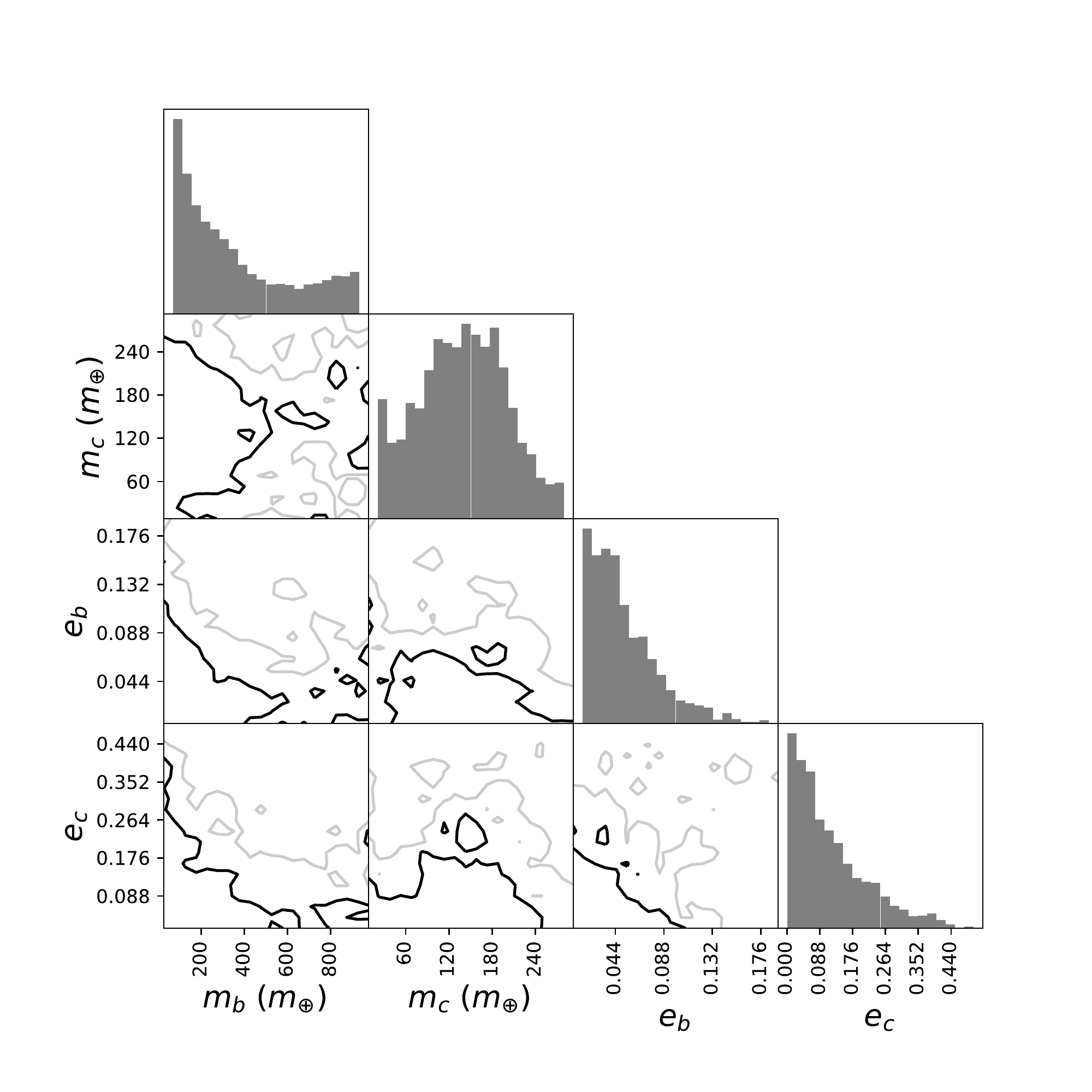}
\caption{Combined posterior mass and eccentricity distributions for the planets in the Kepler-418 system. See Figure \ref{fig4} for description.}
\label{fig12}
\end{figure}

The best-fitting model within the posterior distributions obtained from the DEMCMC, and the observed TTVs for each planet system are shown in Figures \ref{fig3} -- Figure \ref{fig12}. The parameters of the planets are summarized in Table \ref{tab1}. We list our assumed stellar parameters and the planetary parameters we obtain from the fitting procedure. For the best-fitting parameters, we use the median values of the posterior distribution, whereas the reported uncertainties are the standard deviations of the posterior parameter distributions. We confirm that the masses of the planets in Kepler-30 and Kepler-117 systems agree with previous works \citep{Bruno2015,Panichi2017} and we determine that Kepler-302 is a warm Jupiter system. Kepler-487 and Kepler-418 are warm Jupiter candidate systems because their masses are poorly constrained. The companions of the giant planets are several to tens of Earth masses. They are all marked as yellow dots with red circles in Figure \ref{fig1}. Planets in these systems are nearly co-planar (although for Kepler-302 and Kepler-418, $\Delta\Omega$ is not strongly constrained). We further discuss the properties of the systems in the following sections.

\subsubsection{Kepler-30}
Kepler-30 is a typical warm Jupiter system which has attracted considerable attention \citep{fabrycky2012,Sanchis2012,Hadden2017,Panichi2017}. The architecture of Kepler-30 is reminiscent of WASP-47, where, as shown in Figure \ref{fig2}, the Jupiter-sized planet has close-in siblings both inside and outside of its orbit. The orbital separations of planets in Kepler-30, however, are much larger than those found in WASP-47. Moreover, the planets in the Kepler-30 system are on much longer-period orbits. Our system characterization shows that the mutual inclinations between the planets are smaller than $5^\circ$, consistent with the proposition in \citet{Sanchis2012} that Kepler-30 is a co-planar system based on the lack of significant transit duration variations. The planetary masses of Kepler-30 b and Kepler-30 c that we obtain agree with those of \citet{Hadden2017} and \citet{Panichi2017}. The planetary mass of Kepler-30 d is within $2\sigma$ of the value reported by \citet{Panichi2017}, who independently fitted the transit times of Kepler-30. The uncertainties in the Kepler-30 transit times in their results are larger than ours (which were adopted from \citet{Holczer2016}), which may lead to the $2\sigma$ difference in masses that we have found.

\begin{figure}
\vspace{0cm}\hspace{0cm}
\centering
\includegraphics[width=\columnwidth]{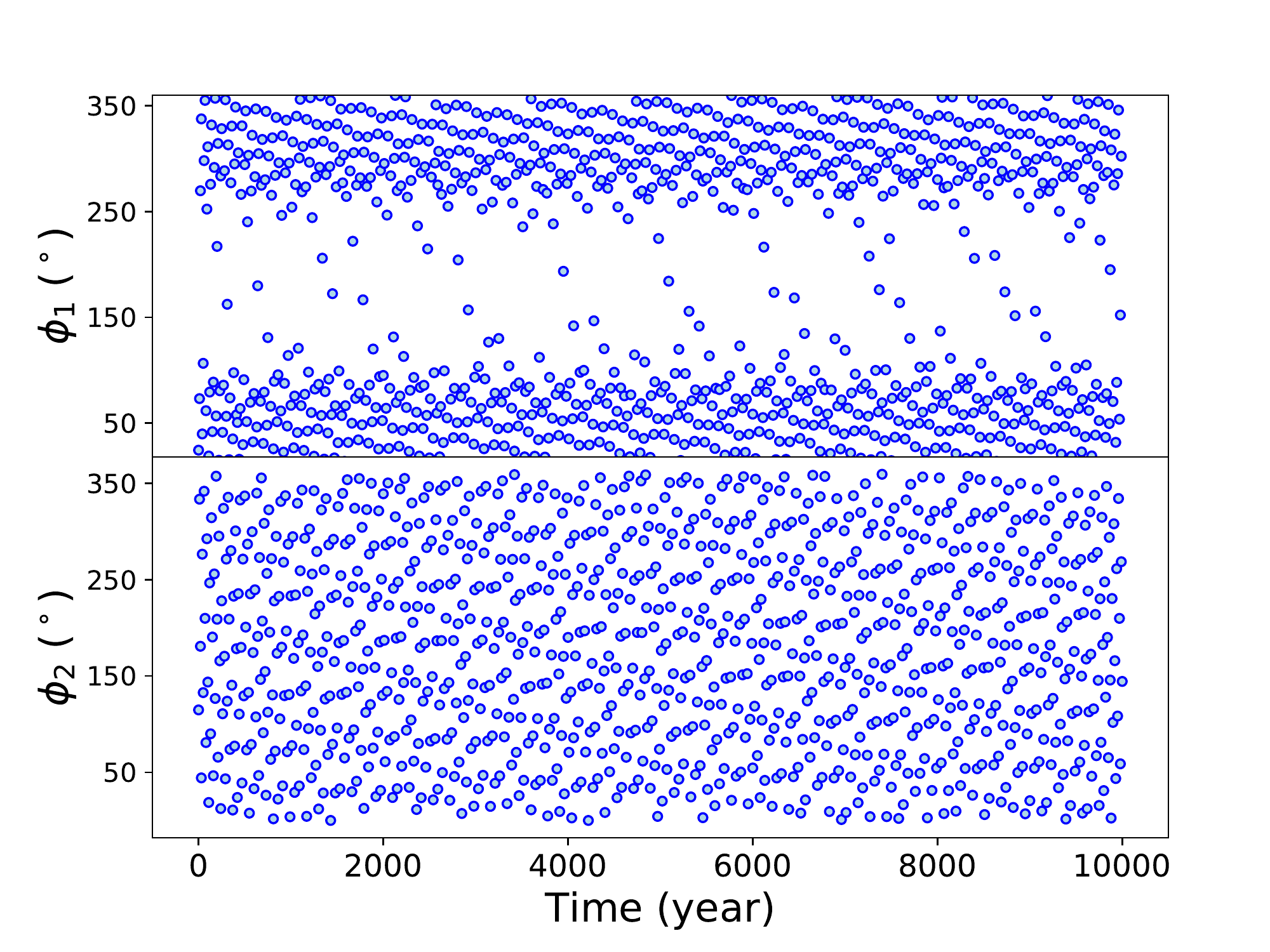}
\caption{The evolution of the resonance angles $\phi_1$ and $\phi_2$ during an evolutionary span of $1\times10^4$ years for Kepler-30 b and Kepler-30 c.}.
\label{fig13}
\end{figure}

Kepler-30 b and Kepler-30 c lie close to the 2:1 orbital commensurability, with an offset of $\Delta=P_{c}/(2P_{b})-1$ $\sim0.032$. We investigate via numerical simulation whether the planet pair is participating in the 2:1 MMR.  We randomly choose 100 sets of planet parameters from the converged posterior distributions and carried out an N-body simulation for each set, each with an integration time of 10000 years. The evolution of the resonance angles  $\phi_1=2\lambda_c-\lambda_b-\varpi_b$ and $\phi_2=2\lambda_c-\lambda_b-\varpi_c$($\varpi=\omega+\Omega$) from one group are shown in Figure \ref{fig13}. The integrations suggest that Kepler-30 b and Kepler-30 c are not in MMR, as the critical arguments for the 2:1 eccentricity-type resonances circulate in all instances. The angle $\phi_1$, however, does spend most of its time around $0^{\circ}$, as shown in Figure \ref{fig13}.

\begin{figure}
\vspace{0cm}\hspace{0cm}
\centering
\includegraphics[width=\columnwidth]{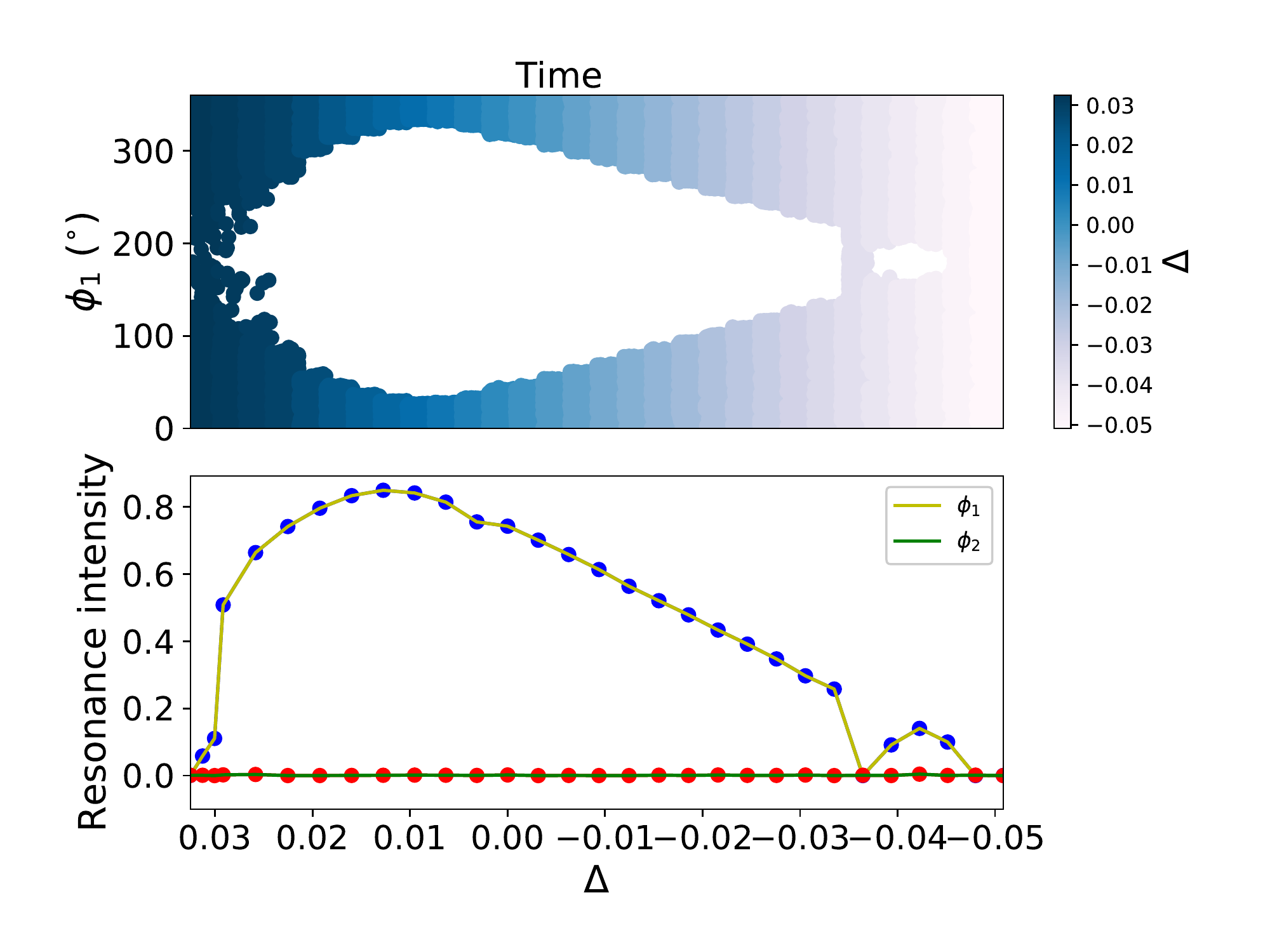}
\caption{\textbf{Upper panel:} Shown from left to right are individual evolutionary traces of the resonance angle $\phi_1=2\lambda_c-\lambda_b-\varpi_b$ over $1\times10^4$ year integrations for different $\Delta$. The figure shows integrations of initial conditions in which the quantity $\Delta$ is varied through adjustment of $P_{c}$. The color represents the MMR offset $\Delta$, and delineates the individual integrations \textbf{Lower panel:} the resonance intensity at different $\Delta$.}.
\label{fig14}
\end{figure}

The closeness of Kepler-30 b and Kepler-30 c to 2:1 MMR is seldom seen in other Kepler planet systems and an indication that the planets probably formed via disk migration. We find that small modifications to either the period or the mass of Kepler-30 c will place the inner pair into resonance. First, we choose the best-fit model and gradually decrease the MMR offset $\Delta$ by adjusting the period of Kepler-30 c. The evolution of $\phi_1$ over $1\times10^4$ year test integrations for different choices of $\Delta$ are shown in the upper panel of Figure \ref{fig14}. We find that when $\Delta$ decreases to 0.030, $\phi_1$ begins to librate. The libration amplitude of $\phi_1$ decreases until $\Delta$ decreases to be near 0, then the libration amplitude begins to increase with the decreases of $\Delta$. Finally, the resonance angle gradually steps out of libration and starts to circulate. We define the resonance intensity as $1-A_{\phi, max}/180$, where $A_{\phi, max}$ is the maximum amplitude (the unit is $^\circ$) of the resonance angle $\phi$ during the $1\times10^4$ evolution. From the lower panel of Figure \ref{fig14} we can see that the resonance intensity increases with the decrease of $\Delta$ at the beginning and decreases to 0 at last. As the required minimal change in $P_c$ to obtain a MMR architecture is beyond the uncertainty of the fitted $P_c$, Kepler-30 b and Kepler-30 c are unlikely to be in MMR. However, they were probably in MMR when they formed and was driven out of MMR by other factors, as many mechanisms \citep{Lithwick2012,Batygin2013,Chatterjee2015} can account for the small change in period ratio. \citet{Panichi2017} analyze the possible formation scenario of Kepler-30 and find that the planets once trapped in a resonant chain can diverge and finally achieve the observed orbital configuration.

\begin{figure}
\vspace{0cm}\hspace{0cm}
\centering
\includegraphics[width=\columnwidth]{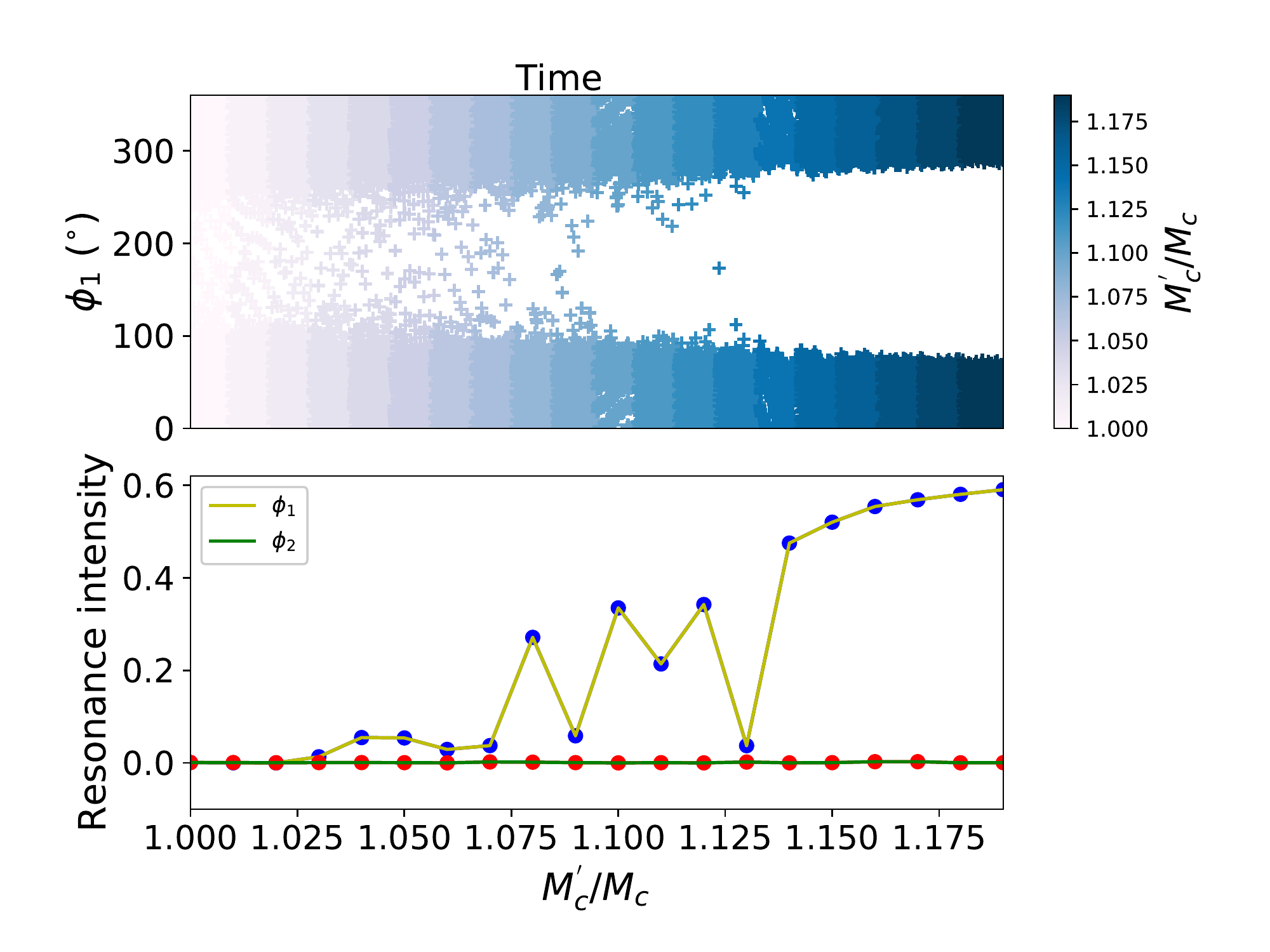}
\caption{\textbf{Upper panel:} Shown from left to right are individual evolutionary traces of the resonance angle $\phi_1=2\lambda_c-\lambda_b-\varpi_b$ over $1\times10^4$ year integrations, with color representing different mass increments for Kepler-30 c ($M_c^{'}/M_c$). \textbf{Lower panel:} the resonance intensity measured as a function of mass increment for Kepler-30 c.}
\label{fig15}
\end{figure}

In addition to changes in the periods, a small modification to the planetary mass can also influence the resonant dynamics of Kepler-30 b and Kepler-30 c. The resonance width increases with the mass of the planets \citep{Deck2013}, therefore, we gradually increase the mass of Kepler-30 c and check if the inner pair display libration in either of the critical arguments. The evolution of the resonance angle $\phi_1$ in the N-body simulation is shown in Figure \ref{fig15}. We find that with increasing planetary mass ($M_c^{'}/M_c$), the resonance angle gradually settles into libration and the resonance intensity also increases from 0 to 0.6. From a dynamical standpoint, the deviation of the inner pair from MMR can be attributed to mass loss experienced by the planets in the past. With planetary periods of tens of days, however, the processes that generate significant planetary mass loss, notably evaporation induced by stellar irradiation, are unlikely to have been important in the Kepler-30 system.

\subsubsection{Kepler-117}

The Kepler-117 system consists of a warm Jupiter in an exterior orbit and a super-Earth on an interior orbit. Our timing analysis finds a mass of  $M_p=31.4{\pm{8.6}}M_{\oplus}$ for Kepler-117 b and a mass $M_p=702{\pm{63}}M_{\oplus}$ for Kepler-117c. Comparing with the previous study of \citet{Bruno2015}, we find agreement within $1\sigma$ when the difference in the assumed stellar mass is taken into account. With periods of $18.78$ and $50.78$ day, the planet pair lies near the 8:3 period commensurability. To check the stability and the potential existence of an 8:3 MMR, we randomly choose 100 converged models from the posterior distribution and carry out numerical $N$-body simulations. We find that all of these system configurations are dynamically stable for 10000$P_2$. The resonance angles $\phi=8\lambda_c-3\lambda_b-i\varpi_b-j\varpi_c$($i+j=5$) always circulate, however, so the planet pair is not in MMR.

\subsubsection{Kepler-302}
In similarity to Kepler-117, Kepler-302's architecture features a warm Jupiter on the outside and a super-Earth on the inside. With limited signal-noise-ratio, however, the uncertainties in the the masses of the planets are large(\textbf{$M_{p,b}=60{\pm{49}}M_{\oplus}$, $M_{p,c}=933\pm{527}M_{\oplus}$}). Interestingly, the eccentricity of the smaller planet Kepler-302 b is as large as $0.198\pm0.065$, which is very different with the eccentricities of other planets in warm Jupiter systems. \citet{Huang2016} proposed that warm Jupiters formed in-situ as warm Jupiters with close-in siblings should have low orbital eccentricities and low mutual inclinations. So Kepler-302 is an exception and needs further study. %The N-body simulation shows that the resonance angles $\phi=17\lambda_c-4\lambda_b-i\varpi_b-j\varpi_c$($i+j=13$) circulate during the integration time of $1\times10^4$ years.

\subsubsection{Kepler-487}
Kepler-487 is classified as a warm Jupiter system in \citet{Huang2016} because of the large planet radius of Kepler-487 b. However, the mass of Kepler-487 b($M_{p,b}=203\pm197 M_{\oplus}$) can not be precisely determined from TTVs, so we consider Kepler-487 to be a warm Jupiter candidate. The planet pair is near 2:5 MMR. We carry out 100 groups of numerical N-body simulation, of which the initial parameters are randomly chosen from the converged chains. The results show that the resonance angles $\phi=5\lambda_c-2\lambda_b-i\varpi_b-j\varpi_c$($i+j=3$) circulate, so the planet pair are not in MMR.

\subsubsection{Kepler-418}
Kepler-418 b was confirmed via the Doppler radial velocity method as a giant planet with an upper mass limit $350$ $M_\oplus$ in two previous studies \citep{Tingley2014,Santerne2016}. The masses calculated from the TTVs($M_{p,b}=241\pm213M_{\oplus}$, $M_{p,c}=138{\pm69}M_{\oplus}$) are poorly constrained (Figure \ref{fig12}) because of the poor signal-noise-ratio. If we constrain the mutual inclination between the two planets to be smaller than $20^{\circ}$ in our fitting procedure, the planetary mass of KOI 1089.02 is $48{\pm41}M_{\oplus}$. With a period ratio $\sim7.1$, we suspect the TTVs of Kepler-418 b and KOI 1089.02 may be not produced via the planetary interaction with each other. With the poorly determined mass of both planets, we did not check this system for mean motion resonances.

\subsection{Giant planet radius anomaly}

The fact that most giant planets in extrasolar systems have larger than-expected radii has been noted since the first transiting hot Jupiers were discovered \citep{Charbonneau2000,Burrows2000,Burrows2004,Gaudi2005}, and is referred to as the radius anomaly. In contrast to Jupiter in the solar system, close-in giant planets in extrasolar systems receive intense stellar irradiation \citep{Spiegel2009,Gaudi2017}, which slows down the cooling rate of the hot Jupiters and results in a larger radius than those isolated Jupiters \citep{Chabrier2004,Spiegel2012}. However, this property alone can not explain all the observed inflated Jupiters. Many other mechanisms are proposed (see \citet{Fortney2010} for a brief review), including occasionally large tidal heating \citep{Jackson2008,Ibgui2009}, enhanced atmospheric opacity \citep{Burrows2007} , semi-convection\citep{Chabrier2007}  and ohmic heating \citep{Batygin2010,Batygin2011,Laughlin2011,Wu2013,Rogers2014,Ginzburg2016}.

\begin{figure}
%\vspace{0cm}\hspace{0cm}
\centering
\includegraphics[width=\columnwidth]{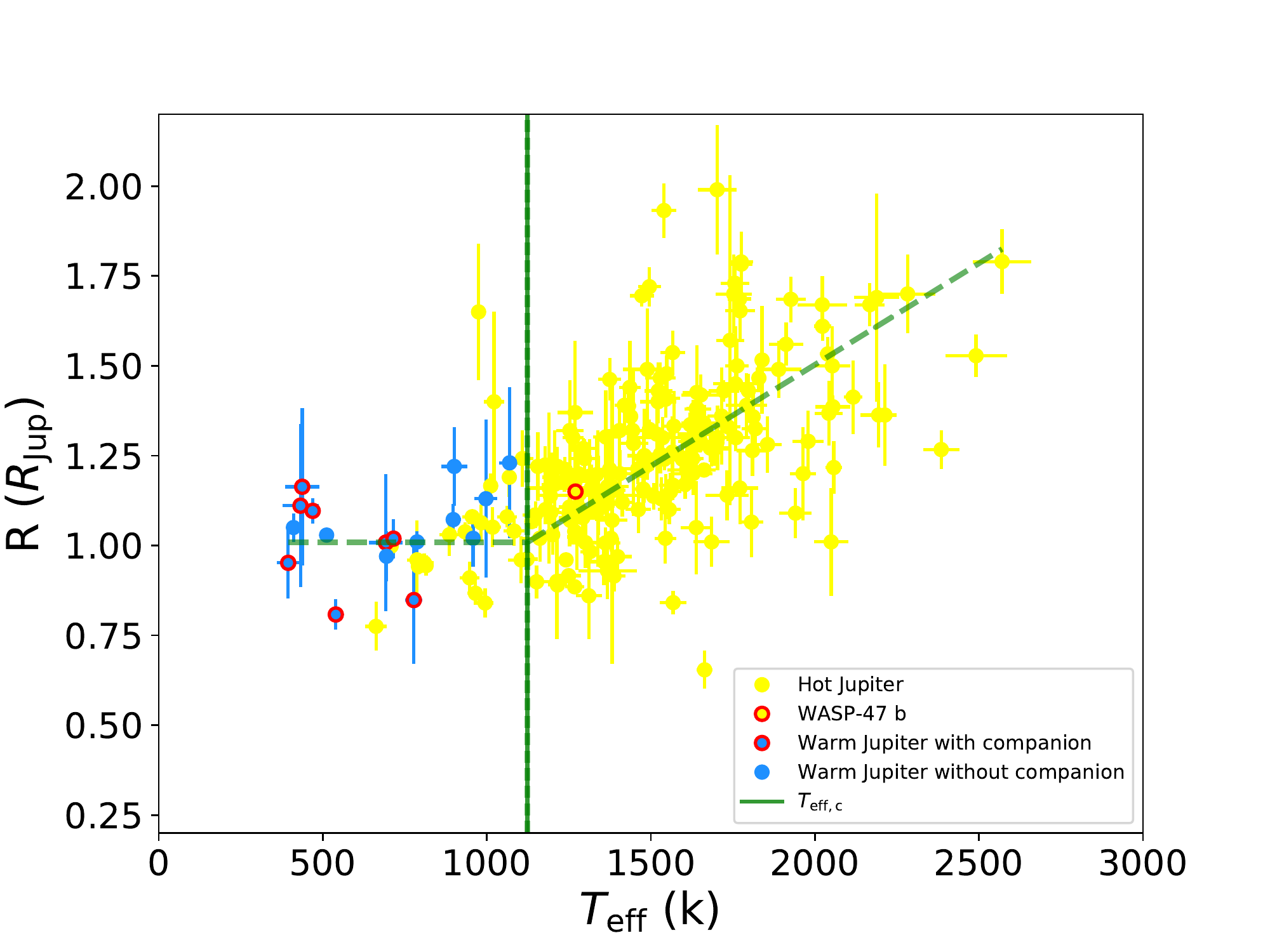}
\caption{Radius vs effective temperature for giant planets ($M_p$ $>$ $0.3M_{\rm Jup}$). Samples of warm Jupiters with companions are from \citet{Huang2016}, other samples are from the Exoplanets.org \citep{Wright2011}. The yellows dots show hot Jupiters ($P<10$ day), while the blue dots show warm Jupiters ($P>10$ day). Dots with red circles represent the planets with companions. The green vertical line represents the critical effective temperature $T_{\rm eff,c}=1123.7\pm3.3$ K obtained by minimizing $\chi^2$ when fitting the $R-T_{\rm eff}$ relation with constant and  linear models, the uncertainties are shown in green vertical dashed lines(As the uncertainties are very small, the lines overlap each other.) We also show the corresponding fitted line models of the R-$T_{\rm eff}$ relationship in green dashed lines.}
\label{fig16}
\end{figure}

In Figure \ref{fig16}, we show the planet radius and effective temperature relation for giant planets with $M_p$ $>$ $0.3M_{\rm Jup}$. Warm Jupiters with known companions are from \citet{Huang2016}, other samples are drawn from Exoplanets.org \citep{Wright2011}. Here we only include warm Jupiters with companions which have mass measurement either in this paper or in other literatures to avoid false alarms such as Kepler-89 d. The warm Jupiters with companions are Kepler-30 c, Kepler-117 c, Kepler-302 c, Kepler-487 b, Kepler-418 b, Kepler-419 b \citep{Dawson2012,Dawson2014}, Kepler-46 b \citep{Nesvorny2012} and Kelper-289 c \citep{Schimitt2014}. The effective temperatures of the planets are calculated according to Eq.1 in \citet{Laughlin2011}. It is clear in Figure \ref{fig16} that planetary radii increase with the effective temperature when  $T_{\rm eff}$ $>$ $1000$ k. When $T_{\rm eff}$ $<$ $1000$ K, however, there is no obvious correlation between the planetary radius and effective temperature. According to ohmic heating \citep{Batygin2010,Batygin2011} , there is a clear tendency toward inflated radius for effective temperature between 1200 K and 1800 K, which gives rise to significant ionization of alkali metals in the atmosphere.

Here we define a critical effective temperature $T_{\rm eff, c}$ to separate Jupiters which show radius inflation and those do not. To simplify the problem, we use two models ($R_1=b_1$,  $R_2=a_2T_{\rm eff}+b_2$) to fit the data on the left and right side of the trial $T_{\rm eff, c}$. $b_1$ is calculated as the weighted mean radius of planets with $T_{\rm eff}\leq T_{\rm eff,c}$, $a_2$ is chosen as $(b_1-b_2)/T_{\rm eff, c}$. So there are two parameters in our fitting, $T_{\rm eff,c}$ and $b_2$. We adopt the MCMC algorithm to minimize $\chi^2=\chi^2(T_{\rm eff}\leq T_{\rm eff,c})+\chi^2(T_{\rm eff}>T_{\rm eff,c})$, and we obtain $T_{\rm eff,c}=1123.7\pm3.3$ K, $b_1=1.00930\pm0.00053$, $b_2=0.3750\pm0.0068$. It is obvious that $T_{\rm eff, c}$(shown as the green vertical line in Figure \ref{fig16}) provides a good separation for Jupiters with companion fraction consistent with zero($T_{\rm eff}>T_{\rm eff,c}$) and those with companion fraction significantly different from zero ($T_{\rm eff}< T_{\rm eff,c}$).  Also, hot Jupiters and warm Jupiters are roughly separated by $T_{\rm eff, c}$.

\subsection{Metallicity}

\begin{figure}
\vspace{0cm}\hspace{0cm}
\centering
\includegraphics[width=\columnwidth]{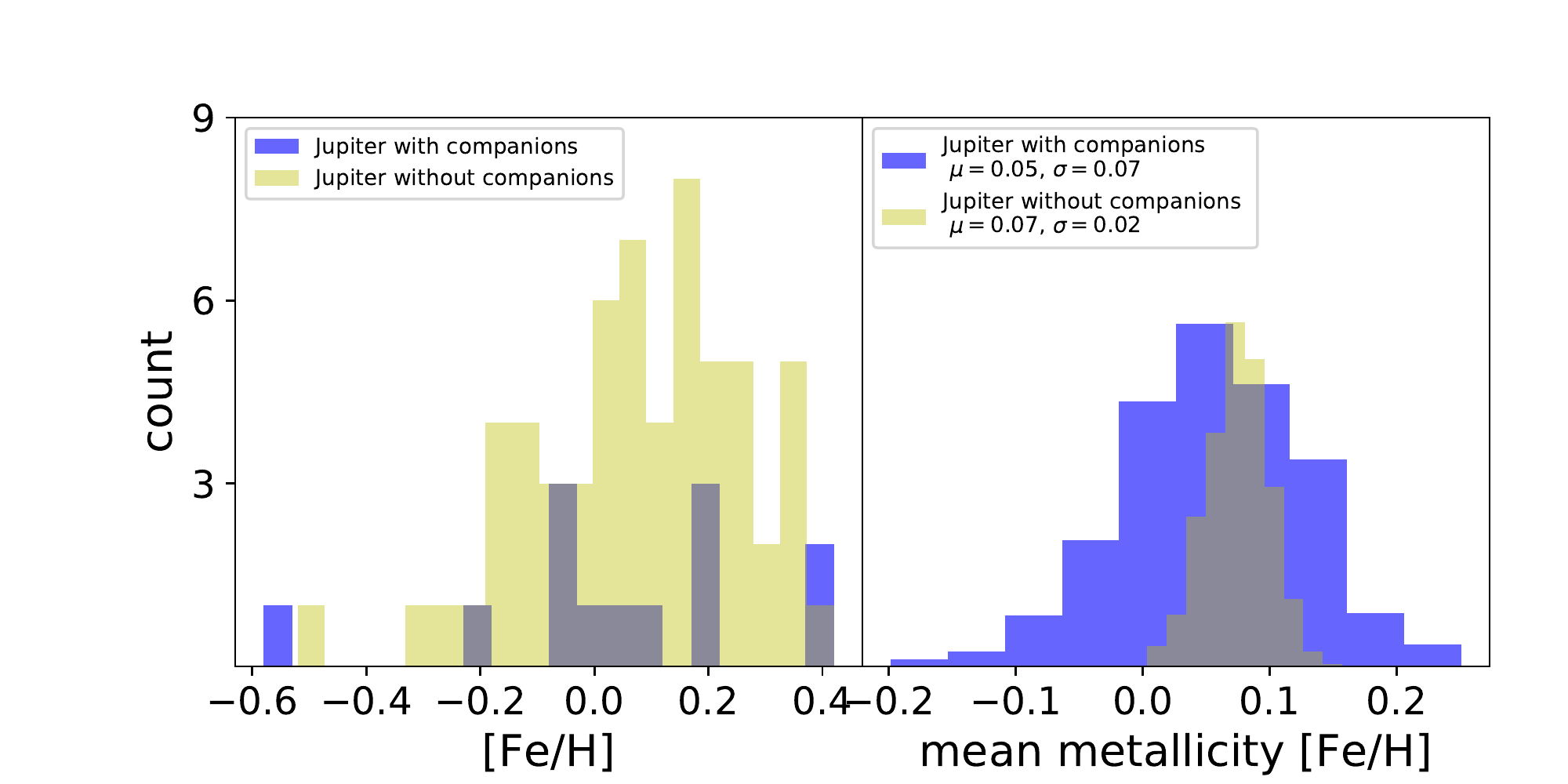}
\caption{Left panel: The stellar metallicity distribution in single-Jupiter systems (yellow) and systems that contain Jupiter-mass planets with additional close-in companions (blue). The single Jupiter samples are from \citet{Huang2016}, the multiple Jupiter samples are from \citet{Huang2016} and the Q1-Q17 DR25 of the NASA Exoplanet Archive. Metallicities for all parent stars are drawn from the Q1-Q17 DR25 of NASA Exoplanet Archive. Right panel: the mean stellar metallicity distribution from a simple bootstrp resampling procedure for single-Jupiter systems(yellow) and systems that contain Jupiter-mass planets with additional close-in companions(blue).}
\label{fig17}
\end{figure}

In Figure \ref{fig17}, we compare the stellar metallicity distributions of single-Jupiter systems and systems that contain Jupiter-mass planets with additional nearby companions.  A companion is deemed ``nearby'' if the period ratio, $0.1$ $<$ $P_{\rm Jup}/P_{\rm comp}$ $<$ $10$, where $P_{\rm Jup}$ and $P_{\rm comp}$ are the orbital periods of the Jupiter and a companion planet, respectively. Single-Jupiter systems are chosen from \citet{Huang2016}, who enforce a strict constraint on the presence of nearby companions. The collection of systems with additional companions was drawn from the Q1-Q17 DR25 of the NASA Exoplanet Archive, and to conform with \citet{Huang2016}, is delineated to have $M_p$ $>$ $0.3M_{\rm Jup}$ or $R_p$ $>$ $0.8R_{\rm Jup}$.  The stellar metallicity distribution of single-Jupiter systems and systems containing Jupiters with additional companions are shown in the left panel of Figure \ref{fig17}. We adopt a simple bootstrap resampling to obtain the mean value and the corresponding uncertainties of the metallicities(shown in the right panel of Figure \ref{fig17}). It is clear from Figure \ref{fig17} that most Jupiter-hosting stars have [Fe/H] $>$ 0, which is consistent with the long-running expectation \citep{Fischer2003,Ida2004} that the formation probability of gas giant planets increases with the metallicity of their host stars. The mean stellar metallicity of single-Jupiter systems has a value of $0.07\pm0.02$, while the mean stellar metallicity of  systems containing Jupiters with additional companions has a value of $0.05\pm0.07$.  The two distributions are statistically indistinguishable here. Future observations such as TESS may provide us more information on the two Jupiter populations to make a better comparison.

 %If the current hint of a metallicity difference between the two populations holds with more data, it would provide suggestive evidence that single Jupiters frequently arise from ``high-$e$'' migration channels. When host star metallicities are higher, there is a larger probability that more than one Jupiter-sized planet forms in a given system, implying that it will be easier to excite dynamical instabilities that leave a single survivor that damps down to a short-period orbit.

\section{Summary and Conclusion}

We have presented the orbital parameters of five warm Jupiter systems in this paper. The planetary systems have been drawn from \citet{Huang2016}, who flagged warm Jupiter systems based on their large planetary radii. We adopted a DE-MCMC-based analysis of these systems using their transit time variations from \citet{Holczer2016}. The results show that only three planets, Kepler-30 c, Kepler-117 c and Kepler-302 c are confirmed to be real warm Jupiters. The masses of the other two-planet systems are poorly constrained because of low signal-noise-ratios, and require future transit or Radial velocity follow-up observations to improve the constraints.

We also conducted numerical simulations of these planetary systems using their best-fit parameters, and found that Kepler-30 b and Kepler-30 c are quite near -- but not in -- low-order mean motion resonances. With only a $10\%$ mass increment or a $0.2\%$ period decrease to Kepler-30 c, the resulting change would put the inner pair into mean motion resonance. Planets in the other warm-Jupiter containing systems are not similarly close to mean motion resonance. 

The isolation of hot Jupiters is often considered as evidence that these planets form via ``high-$e$'' migration \citep{Steffen2012}.  An extension of this mechanism to warm Jupiters, however, is challenged by the observation that many warm Jupiters have nearby companions. The radius-temperature relationship indicates that hot Jupiters and warm Jupiters can be roughly separated by $T_{\rm eff,c}=1123.7\pm3.3$ K, which is obtained as a dividing line between Jupiters which show radius inflation and those do not. Also, $T_{\rm eff,c}$ provides a good separation for Jupiters with companion fraction consistent with zeros and those with companion fraction obviously larger than zero. It is reasonable to suspect that the formation mechanisms for single Jupiter systems and systems which contain Jupiter with companions maybe different. Therefore, we compare the stellar metallicity distribution of the single Jupiter systems and systems containing Jupiter with additional nearby companions, but no statistical conclusions can be made. More data from observations such as TESS, of course, are greatly needed.

%Moreover, the distributions in Figure \ref{fig17} suggests that the stellar metallicity of systems  that  contain  Jupiter-mass  planets with  additional  close-in  companions are slightly smaller than those of single-Jupiter systems. If this relation can be established with additional data, it would suggest that single Jupiters are more likely produced via violent dynamical histories.  In this sense, the formation differences exist not only between hot Jupiter and warm Jupiter systems, but in connection to the degree of isolation. \\

\textbf{Acknowledgments}

We would like to thank the anonymous referee who helped us to improve this paper. This work is supported by the National Natural Science Foundation of China (Grant No. 11503009, 11333002, 11673011, 1166116101), Technology of Space Telescope Detecting Exoplanet and Life” supported by National Defense Science and Engineering Bureau civil spaceflight advanced research project （D030201), and China Scholarship Program. S.W. thanks the Heising-Simons Fundation for their generous support.


\begin{thebibliography}{10}

\bibitem[Agol et al.(2005)]{Agol2005}Agol, E., Steffen, J., \& Sari, R., et al., \ 2005, \mnras, 359, 567
\bibitem[Anderson et al.(2016)]{Anderson2016}Anderson, K.-R., Storch, N.-I., \& Lai, -D., \ 2016, \mnras, 456, 3671
\bibitem[Batygin \& Stevenson(2010)]{Batygin2010}Batygin, K., \& Stevenson, D.-J., \ 2010, \apjl, 714, L238
\bibitem[Batygin et al.(2011)]{Batygin2011}Batygin, K., Stevenson, D.-J., \& Bodenheimer, P.-H., \ 2011, \apj, 738, 1
\bibitem[Batygin \& Morbidelli(2013)]{Batygin2013}Batygin, K. \& Morbidelli, A., \ 2013, \aj, 145, 1
\bibitem[Batygin et al.(2016)]{Batygin2016}Batygin, K., Bodenheimer, P.-H., \& Laughlin, G., \ 2016, \apj, 829, 114
\bibitem[Becker et al.(2015)]{Becker2015}Becker, J.-C., Vanderburg, A., \& Adams, F.-C., et al., \ 2015, \apjl, 812, L18
\bibitem[Bell et al.(1997)]{Bell1997}Bell, K.-R., Cassen, P.-M., Klahr, H.-H., Henning, T., \ 1997, \apj, 486, 372
\bibitem[Beaug{\'e} \& Nesvorn{\'y} (2012)]{Beaug2012}Beaug{\'e}, C., \& Nesvorn{\'y}, D., \ 2012, \apj, 751, 119
\bibitem[Bodenheimer et al.(2000)]{Bodenheimer2000}Bodenheimer, P., Hubickyj, O., \& Lissauer, J.-J., \ 2000, \icarus, 143, 2

\bibitem[Boley et al.(2016)]{Boley2016} Boley, A.~C., Granados Contreras, A.~P., \& Gladman, B.\ 2016, \apjl, 817, L17 



\bibitem[Brooks \& Gelman(1998)]{Brooks1998}Brooks, S.B., Gelman,A.,\ 1998, Journal of Computational and Graphical Statistics, 7:4, 434
\bibitem[Bruno et al.(2015)]{Bruno2015}Bruno, G., Almenara, J.-M., \&, Barros, S.-C.-C., et al., \ 2015, \aap, 573, A124
\bibitem[Burrows et al.(2000)]{Burrows2000}Burrows, A., Guillot, T., \& Hubbard, W.-B., et al., \ 2000, \apjl, 534, L97
\bibitem[Burrows(2004)]{Burrows2004}Burrows, A., Hubeny, I., \& Hubbard, W.-B., et al., \ 2004, \apjl, 610, L53
\bibitem[Burrows et al.(2007)]{Burrows2007}Burrows, A., Hubeny, I., \& Budaj, J., et al., \ 2007, \apj, 661, 502
\bibitem[Charbonneau et al.(2000)]{Charbonneau2000}Charbonneau, D., Brown, T.-M., \& Latham, D.-W., et al., \ 2000, \apjl, 529, L45
\bibitem[Chabrier et al.(2004)]{Chabrier2004}Chabrier, G., Barman, T., \& Baraffe, I., et al., \ 2004, \apjl, 603, L53
\bibitem[Chabrier \& Baraffe(2007)]{Chabrier2007}Chabrier, G., \& Baraffe, I., \ 2007, \apj, 661, L81
\bibitem[Chatterjee et al.(2008)]{Chatterjee2008}Chatterjee, S., Ford, E.-B., Matsumura, S., \& Rasio, F.-A., \ 2008, \apj, 686, 580
\bibitem[Chatterjee \& Ford(2015)]{Chatterjee2015}Chatterjee, S. \& Ford, E.-B., \ 2015, \apj, 803, 33
\bibitem[Chen et al.(2013)]{Chen2013}Chen, Y.-Y., Liu, H.-G., Zhao, G., \& Zhou, J.-L., \ 2013, \apj, 769, 26
\bibitem[Chiang \& Laughlin(2013)]{Chiang2013}Chiang, E., \& Laughlin, G., \ 2013, \mnras, 431, 3444
\bibitem[Dawson et al.(2012)]{Dawson2012}Dawson, R.-T., Johnson, J.-A., \& Morton, T.-G., et al., \ 2012, \apj, 761, 163
\bibitem[Dawson et al.(2014)]{Dawson2014}Dawson, R.-T., Johnson, J.-A., \& Fabrycky, D.-C., et al., \ 2014, \apj, 791, 89
\bibitem[Deck et al.(2013)]{Deck2013}Deck, K.-M., Payne, M., \& Holman, M.-J., \ 2013, \apj, 774, 129
\bibitem[Deck et al.(2014)]{Deck2014} Deck, K.~M., Agol, E., Holman, M.~J., \& Nesvorn{\'y}, D.\ 2014, \apj, 787, 132 
\bibitem[Dong et al.(2014)]{Dong2014}Dong, S., Katz, B., \& Socrates, A., \ 2014, \apjl, 781, L5
\bibitem[Fabrycky et al.(2012)]{fabrycky2012}Fabrycky, D.-C., Ford, E.-B., \& Steffen, J.-H., et al., \ 2012, \apj, 750, 114
\bibitem[Fang \& Margot(2012)]{fang2012}Fang, J., \& Margot, J.-L., \ 2012, \apj, 761, 92
\bibitem[Fischer \& Valenti(2003)]{Fischer2003}Fischer, D.-A., \& Valenti, J.-A., \ 2003, ASPCS, 294, 117
\bibitem[Ford(2006)]{Ford2006}Ford, E.~B.,\ 2006, \apj, 642,505 
\bibitem[Fortney \& Nettelmann(2010)]{Fortney2010}Fortney, J.-J., \& Nettelmann, N., \ 2010, \ssr, 152, 423
\bibitem[Gaudi(2005)]{Gaudi2005}Gaudi, B.-S., \ 2005, \apjl, 628, L73
\bibitem[Goldreich \& Tremaine(1980)]{Goldreich1980}Goldreich, P., \& Tremaine, S., \ 1980, \apj, 241, 425
\bibitem[Gaudi et al.(2017)]{Gaudi2017}Gaudi, B.-S., Stassun, K.-G., \& Collins, K.-A., et al., \ 2017, \nat 546, 514
\bibitem[Ginzburg \& Sari(2016)]{Ginzburg2016}Ginzburg, S., \& Sari, R., \ 2016, \apj, 819, 116
\bibitem[Hadden \& Lithwick(2017)]{Hadden2017}Hadden, S. \& Lithwick, Y., \ 2017, \aj, 154, 5
\bibitem[Hellier et al.(2012)]{Hellier2012}Hellier, C., Anderson, D.-R., \& Collier Cameron, A., et al., \ 2012 \mnras, 426, 739
\bibitem[Holczer et al.(2016)]{Holczer2016} Holczer, T., Mazeh, T., Nachmani, G., Jontof-Hutter, D., Ford, E. B., Fabrycky, D., Ragozzine, D., Kane, M., \& Steffen, J. H. 2016, \apjs, 225, 9
\bibitem[Holman \& Murray(2005)]{Holman2005}Holman, M.-J. \& Murray, N.-W., \ 2005, Science, 307, 1288
\bibitem[Huang et al.(2016)]{Huang2016}Huang, C., Wu, Y., \& Triaud, A.-H.-M.-J., \ 2016, \apj, 825, 98
\bibitem[Ibgui \& Burrows(2009)]{Ibgui2009}Ibgui, L., \& Burrows, A., \ 2009, \apj, 700, 1921
\bibitem[Ida \& Lin(2004)]{Ida2004}Ida, S., \& Lin, D.-N.-C., \ 2004, \apj, 616, 567
\bibitem[Ida \& Lin(2010)]{Ida2010}Ida, S., \& Lin, D.-N.-C., \ 2010, \apj, 719, 810
\bibitem[Jackson(2008)]{Jackson2008}Jackson, B., Greenberg, R., \& Barnes, R., \ 2008, \apj, 681, 1631
%\bibitem[Johnson et al.(2017)]{Johnson2017} Johnson, J.~A. \& Petigura, E.~A., et al. \ 2017, arXiv:1703.10402v2 
\bibitem[Jontof-Hutter et al.(2016)]{JH2016}Jontof-Hutter, D., Ford, E.-B., \& Rowe, J.-F., et al., \ 2016, \apj, 820, 39
\bibitem[Kozai(1962)]{kozai1962}Kozai, Y., \ 1962, \aj, 67, 591
\bibitem[Laughlin et al.(2011)]{Laughlin2011}Laughlin, G., Crismani, M., \& Adams, F.-C., \ 2011, \apjl, 729, L7
\bibitem[Lee et al.(2014)]{Lee2014}Lee, E.-J., Chiang, E., \& Ormel, C.-W., \ 2014, \apj, 797, 95
\bibitem[Lin \& Papaloizou(1986)]{Lin1986}Lin, D.-N.-C., \& Papaloizou, J., \ 1986, \apj, 309, 846
\bibitem[Lin et al.(1996)]{Lin1996}Lin,D.-N.-C., Bodenheimer, P., \& Richardson, D.-C., \ 1996, \nat, 380, 606
\bibitem[Lissauer et al.(2011)]{lissauer2011}Lissauer, J.-J., Ragozzine, D., \& Fabrycky, D.-C., et al., \ 2011, \apjs, 197 8
\bibitem[Lithwick \& Wu(2012)]{Lithwick2012}Lithwick, Y., \& Wu, Y., \ 2012, \apjl, 756, L11
\bibitem[Lithwick et al.(2012)]{Lithwick-12012}Lithwick, Y., Xie, J., \& Wu, Y., \ 2012, \apj, 761, 122
\bibitem[Mart{\'i} \& Beaug{\'e}(2015)]{Mart2015}Mart{\'i}, J.-G., \& Beaug{\'e}, C., \ 2015, International Journal of Astrobiology, 14, 313
\bibitem[Masset \& Papaloizou(2003)]{Masset2003}Masset, F.-S., \& Papaloizou, J.-C.-B., \ 2003, \apj, 588, 494
\bibitem[Mullally et al.(2015)]{Mullally2015}Mullally, F., Coughlin, J.-L, \& Thompson, S.-E., et al., \ 2015, \apjs, 217, 31
\bibitem[Nagasawa et al.(2008)]{Nagasawa2008}Nagasawa, M., Ida, S. \& Bessho, T., \ 2008, \apj, 678, 498
\bibitem[Naoz et al.(2011)]{Naoz2011}Naoz, S., Farr, W.-M., Lithwick, Y. Rasio, F.-A., \& Teyssandier, J., \ 2011, \nat, 473, 187
\bibitem[Naoz et al.(2012)]{Naoz2012}Naoz, S., Farr, W.-M., \& Rasio, F.-A., \ 2012, \apj, 754, L36
\bibitem[Nelson et al.(2014)]{Nelson2014}Nelson, B., Ford, E.-B., \& Payne, M.-J., \ 2014, \apjs, 210, 11
\bibitem[Nesvorn{\'y} et al.(2012)]{Nesvorny2012}Nesvorn{\'y}, D., Kipping, D.-M., \& Buchhave. L.-A., et al., \ 2012, Science, 336, 1133
\bibitem[Neveu-VanMalle et al.(2016)]{Neveu2016}Neveu-VanMalle, M., Queloz, D., \& Anderson, D.-R., et al., \ 2016, \aap, 586, A93 
\bibitem[Panichi et al.(2017)]{Panichi2017}Panichi, F., Go{\'z}dziewski, K., \& Migaszewski, C., et al., \ 2017, arXiv: 1707.04962
\bibitem[Rasio \& Ford(1996)]{Rasio1996}Rasio, F.-A., \& Ford, E.-B., \ 1996, \apj, 274, 954
\bibitem[Rogers \& Komacek(2014)]{Rogers2014}Rogers, T.-M., \& Komacek, T.-D., \ 2014, \apj, 794, 132
\bibitem[Sanchis-Ojeda et al.(2012)]{Sanchis2012}Sanchis-Ojeda, R., Fabrycky, D.-C., \& Winn, J.-N., et al., \ 2012, \nat, 487, 449
\bibitem[Santerne et al.(2016)]{Santerne2016}Santerne, A., Moutou, C., \& Tsantaki, M., et al., \ 2016, \aap, 587, A64
\bibitem[Schmitt et al.(2014)]{Schimitt2014}Schmitt, J.-R., Agol, E., \& Deck, K.-M., et al., \ 2014, \apj, 795, 167
\bibitem[Shu et al.(1994)]{Shu1994}Shu, F., Najita, J., \& Ostriker, E., et al., \ 1994, \apj, 429, 781
\bibitem[Sinukoff et al.(2017)]{Sinukoff2017}Sinukoff, E., Howard, A.-W., \& Petigura, E.-A., et al., \ 2017, \aj, 153, 70
\bibitem[Spiegel et al.(2009)]{Spiegel2009}Spiegel, D.-S., Silverio, K., \& Burrows, A., \ 2009, \apj, 699, 1487
\bibitem[Spiegel \& Burrows(2012)]{Spiegel2012}Spiegel, D.S., \& Burrows, A.,\  2012, \apj, 745, 174
\bibitem[Steffen et al.(2012)]{Steffen2012}Steffen, J.-H., Ragozzine, D., Fabrycky, D.-C., et al., \ 2012, PNAS, 109, 7982
\bibitem[Tingley et al.(2014)]{Tingley2014}Tingley, B., Parviainen, H., \& Gandolfi, D., et al., \ 2014, \aap, 567, A14
\bibitem[Ter Braak(2006)]{Ter2006} Ter Braak, C.J.F.\ 2006, Stat Comput, 16, 239
\bibitem[Wang et al.(2017)]{Wang2017}Wang,S., Wang, Y., \& Zhang, X., \ 2017, Submitted.
\bibitem[Wang et al.(2017)]{Wangwu2017}Wang, S., Wu, D.-H., \& Addison, B.-C.,et al., \ 2018, \aj, 155, 73
\bibitem[Wang et al.(2017)]{Yong2017}Wang, Y.-H., Wang, S., \& Liu, H.-G., et al.,\ 2017, \aj, 154,49
\bibitem[Weiss et al.(2017)]{Weiss2017}Weiss, L.M., Deck, K.-M., \& Sinukoff, E., et al., \ 2017, \aj, 153, 265
\bibitem[Wu et al.(2007)]{Wu2007}Wu,Y., Murray, N.-W., \& Ramsahai,J.-M., \ 2007, \apj, 670, 820
\bibitem[Wu \& Lithwick(2013)]{Wu2013}Wu. Y., \& Lithwick, Y., \ 2013, \apj, 763, 13
\bibitem[Wright et al.(2011)]{Wright2011}Wright, J.-T., Fakhouri, O., \& Marcy, G.-W., et al., \ 2011, \pasp, 123, 412
\bibitem[Wu \& Lithwick(2011)]{Wu2011}Wu, Y., \& Lithwick, Y., \ 2011, \apj, 735, 109
\bibitem[Xie (2014)]{Xie2014}Xie, J.-W., \ 2014, \apj, 786, 153
\bibitem[Xie et al.(2016)]{Xie2016}Xie, J.-W., Dong, S., \& Zhu, Z., et al., \ 2016, PNAS, 113, 11431
\bibitem[Zakamska \& Tremaine (2004)]{Zakamska2004}Zakamska, N.-L., \& Tremaine, S., \ 2004, \aj, 128, 869


%X
%Y
%Z
\end{thebibliography}
\end{document}